\documentclass[aps,pra,10pt,twocolumn,amsfonts,superscriptaddress,showpacs,floatfix, notitlepage]{revtex4-1}

\usepackage{braket,verbatim,bm,bbold,amsmath,amssymb,epsfig,float,color}
\usepackage[colorlinks,bookmarks=false,citecolor=blue,linkcolor=blue,hyperfootnotes=true,urlcolor=blue]{hyperref}
\usepackage{tikz}
\usepackage{tikz-network}
\usetikzlibrary{patterns,decorations.pathreplacing}
\definecolor{myred}{RGB}{232,102,102}

\newcommand{\bw}{\begin{widetext}}
\newcommand{\ew}{\end{widetext}}
\newcommand{\be}{\begin{equation}}
\newcommand{\ee}{\end{equation}}
\newcommand{\bea}{\begin{eqnarray}}
\newcommand{\eea}{\end{eqnarray}}

\def\nn{\nonumber\\}

\definecolor{violet}{rgb}{0.62,0,1}
\definecolor{lightblue}{rgb}{0.12,0.56,1}
\definecolor{green}{rgb}{0.13,0.55,0.13}

\begin{document}

\title{Prethermalisation and Thermalisation in the Entanglement Dynamics}  

\author{Bruno Bertini}
\affiliation{Department of Physics, Faculty of Mathematics and Physics,
University of Ljubljana, Jadranska 19, SI-1000 Ljubljana, Slovenia}

\author{Pasquale Calabrese}
\affiliation{SISSA and INFN, Sezione di Trieste, via Bonomea 265, I-34136, Trieste, Italy}
\affiliation{International Centre for Theoretical Physics (ICTP), I-34151, Trieste, Italy}

\date{\today}

\begin{abstract}
{We investigate the crossover of the entanglement entropy towards its thermal value in nearly integrable systems.} We employ equation of motion techniques to study the entanglement dynamics in a lattice model of weakly interacting spinless fermions after a quantum quench. For weak enough interactions we observe a two-step relaxation of the entanglement entropies of finite subsystems. Initially the entropies follow a nearly integrable evolution, approaching the value predicted by the Generalized Gibbs Ensemble (GGE) of the unperturbed model. Then, they start a slow drift towards the thermal stationary value described by a standard Gibbs Ensemble (GE). While the initial relaxation to the GGE is independent of the interaction, the slow drift from GGE to GE values happens on time scales proportional to the inverse interaction squared. For asymptotically large times and subsystem sizes the dynamics of the entropies can be predicted using a modified quasiparticle picture that keeps track of the evolution of the fermionic occupations caused by the integrability breaking. This picture gives a quantitative description of the results as long as the integrability-breaking timescale is much larger than the one associated with the (quasi) saturation to the GGE. In the opposite limit the quasiparticle picture still provides the correct late-time behaviour, but it underestimates the initial slope of the entanglement entropy.    
\end{abstract}

\maketitle

The non-equilibrium dynamics of the entanglement in many-body systems is currently attracting huge attention, effectively bridging the gap between condensed matter, quantum information, and high-energy physics. Some of the big questions in this context concern the onset of thermalisation in isolated many-body systems~\cite{PolkonikovRMP11, GE15, DKPR15, SI}, the origin of thermodynamic entropy~\cite{C:18, DLS:13, BAH:15, G:14, SPR:11,ckc-14,kaufman-2016}, the scrambling of quantum information in quantum chaotic systems~\cite{ABGH:15, HRT:15, AAAL:10, HM:13, LS:14, CLM:16, HP:07, SS:08, LM:CFT,AC:19,mac-20}, as well as the simulability of the quantum many-body dynamics via classical computers~\cite{SWVC:PRL, SWVC:NJP, PV:08, HCTDL:12, D:17}.

A fascinating aspect of this problem lies in its universality: the entanglement dynamics does not seem to depend much on the microscopic details of the many-body system. For instance, considering a quantum quench from a separable state one typically observes linear growth of the entanglement followed by saturation. When first observed in the context of $(1+1)$-dimensional conformal field theory (CFT), this phenomenon has been explained assuming that the entanglement is transported by pairs of correlated quasiparticles~\cite{CC}. This intuitive {\it quasiparticle picture} can be used in systems with stable quasiparticle excitations, such as 
free~\cite{FC:exactXY, ep-08, nr-14, bkc-14, bhy-17, hbmr-17, buyskikh-2016, cotler-2016, triplets, intertwined, BFPC18} and interacting~\cite{CA, CA2, alba-inh, ABF19, mkz-17, mbpc-17} integrable models, but it does
not account for the fact that the same qualitative behaviour is also observed in systems with no detectable quasiparticle content such as holographic CFTs~\cite{ABGH:15, LM:CFT, LS:14} or generic interacting systems~\cite{ckt-18, FNR:longrangehigherd, PL:kickedIsing, KH:NonIntEnt, lauchli-2008, PBCP20, GoLa19, BKP:entropy,smg-20,brydges-2018} --- in essence, the only known cases where the entanglement does not behave as described above are connected with localisation~\cite{NH:MBL, ZPP:08, DMCF:06,bpm-12,Vosk2014, ModulatedIsing, ModulatedIsing2}, quenched disorder~\cite{NRH:18,isl-12}, confinement~\cite{KCTC:17,lsmpcg,jkr-19,clsv-20, Schwinger}, or presence of quantum scars~\cite{scar1,scar2}. 

Recently, an alternative explanation for the universality of the entanglement dynamics has arisen by studying the so-called (local) random unitary circuits~\cite{NRVH:17}, where the dynamics is completely random in space and the only constraint is given by the locality of interactions. In this case, one quantifies the amount of entanglement between two portions of the system by measuring the surface of the minimal space-time membrane separating them. This {\it minimal membrane picture} has been analytically tested in random unitary circuits~\cite{Nahum:operatorspreadingRU} and it is believed to describe, at least qualitatively, the entanglement spreading in generic (non-integrable) systems in any spatial dimension (see Ref.~\cite{ZN:nonrandommembrane} for quantitative comparisons). 

The two pictures discussed above rely on very different physical mechanisms and in general give different predictions. For instance, their predictions for the dynamics of the entanglement of disjoint regions~\cite{ABGH:15, AC:19} or that of a connected region in finite volume~\cite{NRVH:17, BKP:entropy,mac-20} are qualitatively different. {In essence, while correlated quasiparticles produce disentanglement whenever the pairs find themselves in the same subsystem in the course of the evolution, no disentanglement is observed in the non-integrable case.} A natural question is then what happens to the entanglement dynamics when the integrability is broken only weakly. In this case one would expect the two different mechanisms underlying the above picture to somehow coexist until the metastable quasiparticles decay. 

Weakly non-integrable systems are per se very interesting. Indeed, recent theoretical~\cite{AFpret, RFGpret, MK:prethermalization, RoschPRL08, KollarPRB11, worm13, MarcuzziPRL13, EsslerPRB14, NIC14, Fagotti14, konik14, BF15, CTGM:pret, knap15, SmacchiaPRB15, BEGR:PRL, BEGR:PRB, FC15, MenegozJStatMech15, KaminishiNatPhys15, D-14,dv-17, DeRoeck, FGV:19, DBD:20, LWGV:20} and experimental~\cite{gring-2012,LangenReview, Dysprosiumcradle} investigations pointed out that these systems display crossovers from integrable to non-integrable dynamics that are reminiscent of those described by the celebrated Kolmogorov-Arnold-Moser theory in few-particle classical integrable systems. The intuitive picture is that approximate conservation laws in the system generate a \emph{separation of time-scales}~\cite{DeRoeck}. A symmetry-breaking term of order $U$ becomes effective over a time-scale that increases with $1/{U}$. For small enough $U$ this is much larger than the relaxation time and the system relaxes as if the approximate conservation law were exact~\cite{EsslerPRB14}. At later times the symmetry breaking becomes effective and observables drift towards the true equilibrium state~\cite{BEGR:PRL}. This phenomenon, dubbed {\it prethermalization}~\cite{MK:prethermalization}, is of crucial practical importance: it shows that integrability --- although fragile --- can be {\it dynamically robust} and hence \emph{observable}. This is the ultimate explanation of why many cold-atom experiments detect traces of integrable many-body dynamics~\cite{coldatoms, Newtoncradle,sbdd-20,langen-2015}.

A major obstacle is that, aside from being physically very rich, the {\it prethermalization regime} is also very hard to access. One needs to follow the out-of-equilibrium dynamics of the (strictly speaking non-integrable) system for very long-times and the methods (both analytical and computational) to do that are very scarce. One might try to access this regime in some special class of non-integrable systems, like the recently discovered dual-unitary quantum circuits~\cite{BKP:dualunitary}, which led to the only available exact results on the entanglement dynamics in locally-interacting non-integrable systems~\cite{PBCP20, BKP:entropy, GoLa19}. However, even though these systems allow to study the weakly non-integrable regime, the aforementioned results on the entanglement dynamics turn out to be independent of the integrability breaking. 
In this paper, we follow an alternative route and address this question focusing on what is arguably the simplest non-trivial setting. We consider a system of weakly interacting fermions on the lattice, which we analyse by means of Equation of Motion techniques~\cite{SK:EOM,INJPhys,NessiArxiv15,BEGR:PRL,ESY:QBE,LS:QBE, BEGR:PRL, BEGR:PRB, Bonitzbook, FMS:NIQBE, FMS:QBE, KBbook}. This is an approximate method based on the truncation of the infinite hierarchy of evolution  equations for connected fermionic cumulants of increasing size (equivalent to the BBGKY hierarchy for reduced density operators~\cite{BEGR:PRB, Bonitzbook}). {Our main finding is the identification of a regime --- integrability breaking much smaller than inverse subsystem size --- where the entanglement dynamics is quantitatively described for all times by a modified quasiparticle picture in which the contribution of each pair to the entanglement is (slowly) time-dependent.}

The rest of this paper is laid out as follows. In Section~\ref{sec:model} we introduce the model and its basic properties. In Section~\ref{sec:prob} we describe the setting considered and briefly recall some basic facts about Equations of Motion. In Sec~\ref{sec:EE} we discuss our results for the dynamics of the the entanglement, and, in Sec.~\ref{sec:quasiparticles}, we interpret them in terms of a modified quasiparticle picture. Finally in Sec.~\ref{sec:conclusions} we report our conclusions.   

\section{The model}
\label{sec:model}

We consider a system of weakly interacting spinless fermions on a one-dimensional lattice of length $L$ whose dynamics are described by the following Hamiltonian
\begin{align}
H[J_2,\delta,U] &= H_1+H_2+H_U, \label{Eq:Ham} \\
H_1&=-J_1\sum_{l=1}^L \Big[1+(-1)^l\delta \Big] \Bigl( c^\dag_l c^{\phantom\dagger}_{l+1}+\textrm{H.c.}\Bigr), \nonumber\\ 
H_2&=-J_2\sum_{l=1}^L  \Bigl[ c^\dag_l c^{\phantom\dagger}_{l+2}+\textrm{H.c.}\Bigr],\nonumber \\
H_U&={U}\sum_{l=1}^L n_ln_{l+1} \nonumber\,.
\end{align}
Here  $c^\dag_i$ and $c^{\phantom{\dag}}_i$ are respectively fermionic creation and annihilation operators obeying the canonical anticommutation relations
\be
\{c^\dag_i,c^{\phantom{\dag}}_j\}=\delta_{i,j}\ ,\qquad
\{c_i,c_j\}=0\,,
\ee
and we imposed periodic boundary conditions $c^{\phantom\dagger}_{L+1}\equiv c^{\phantom\dagger}_{1}$. For definiteness, from now on we consider the length $L$ to be even and set $J_1=1$.  

\begin{figure*}
\begin{tabular}{l}
\includegraphics[width=0.425\textwidth]{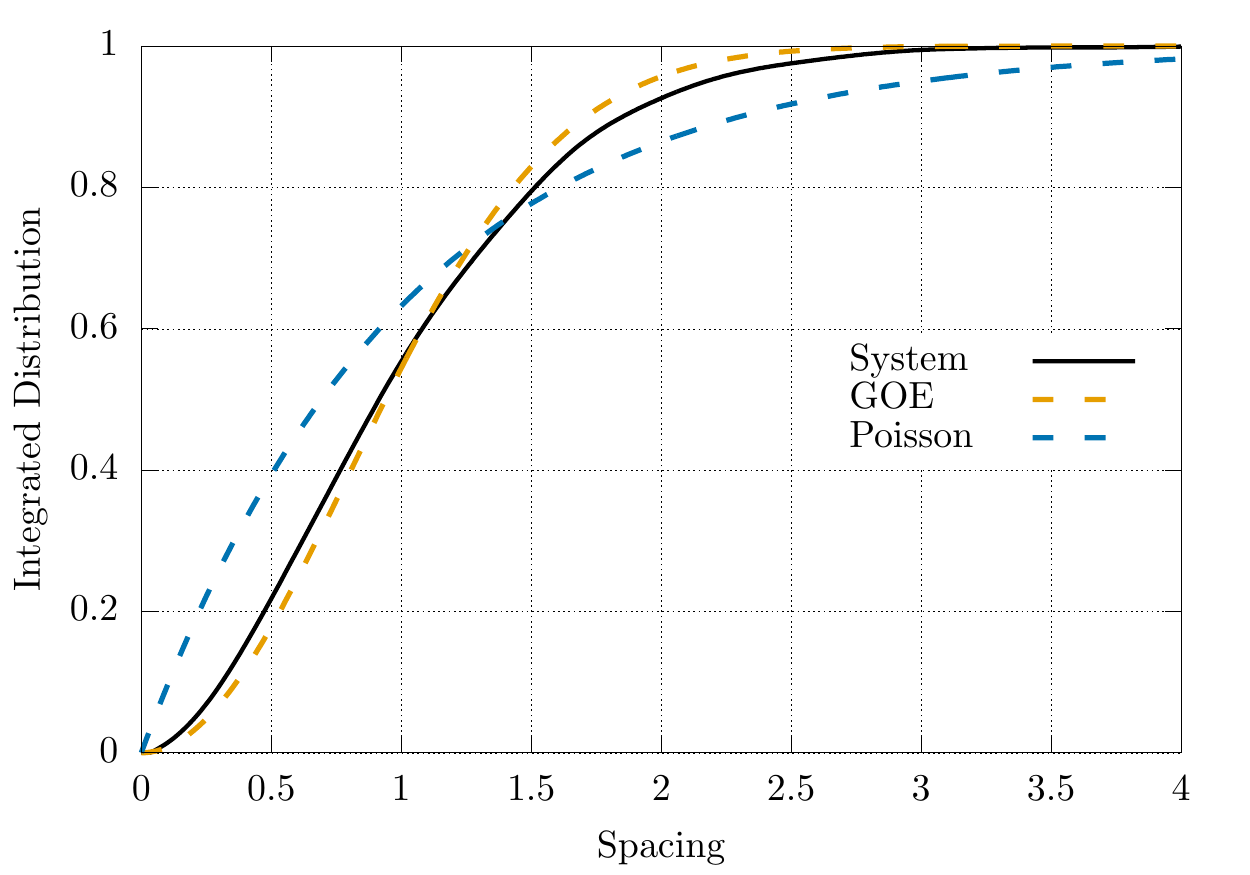}\qquad \includegraphics[width=0.425\textwidth]{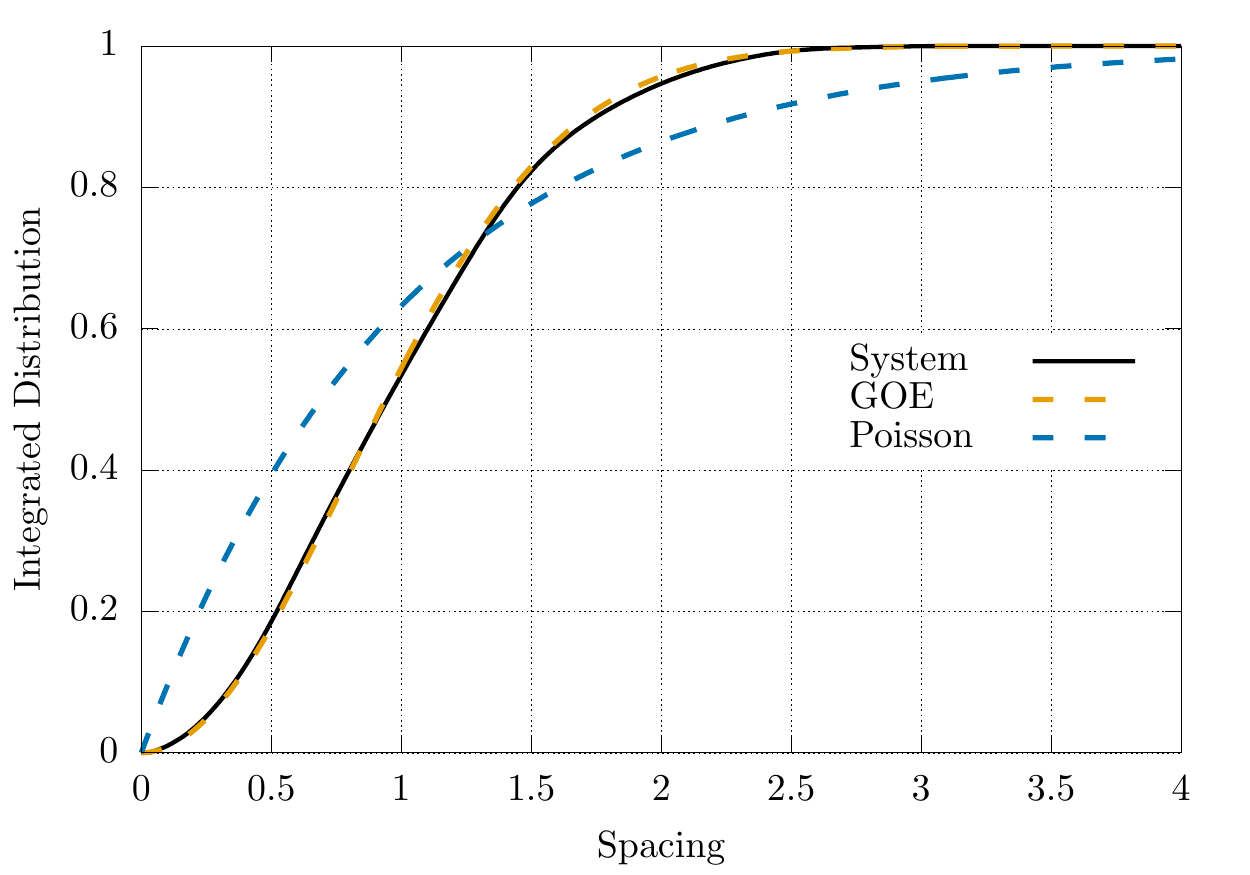}
\end{tabular}
\caption{Cumulative (integrated) level-spacing distributions of \eqref{Eq:Ham} for $L=16$ compared with Poisson and Wigner GOE distributions. Left panel: $\delta=0.5$, $J_2=0$, $U=1$. Right panel: $\delta=0.5$, $J_2=0.5$, $U=1$.} 
\label{Fig:Levelspacing}
\end{figure*}

As discussed in Refs.~\cite{BEGR:PRL, BEGR:PRB} the Hamiltonian is integrable for ${U=0}$ --- where it describes free fermions --- and for ${\delta = J_2 = 0}$ --- where it can be mapped into a XXZ spin-1/2 chain in an external magnetic field \cite{Orbach} through a Jordan-Wigner transformation. Moreover for $J_2=0$ and $\delta,U\ll1$ its low energy description is given by the quantum sine-Gordon model~\cite{EKreview}.  Away from these points $H[J_2,\delta,U]$ is believed to be non-integrable. This is confirmed the statistics of the level spacing of its unfolded spectrum. As shown in Fig.~\ref{Fig:Levelspacing}, level spacings are well described by the Gaussian Orthogonal Ensemble of random matrices.

Since we are interested in the regime of small interactions, it is convenient to diagonalise the quadratic part of the Hamiltonian. This is achieved by the following linear mapping (see the Supplemental Material of Ref.~\cite{BEGR:PRL}) 
\be
c_l=\frac{1}{\sqrt{L}}\sum_{k>0}\sum_{\eta=\pm}
\gamma_\eta(l,k|\delta)\alpha_\eta(k)\ .
\label{Eq:BogoTrans}
\ee
Here the sum runs over $k\in 2\pi/L\, \mathbb Z_{L/2}\subset[0,\pi]$ and 
\be
\begin{split}
&\gamma_{\pm}(2j-1,k|\delta)=e^{-i k (2j-1)}\,,\\
&\gamma_{\pm}(2j,k|\delta)=\pm e^{-i k 2 j} e^{-i \varphi_k(\delta)}\,,\\ 
&e^{-i \varphi_k(\delta)}=\frac{-\cos k+ i \delta\sin k}{\sqrt{\cos^2 k+ \delta^2\sin^2k}}~.
\end{split}
\ee
This transformation is a combination of a two-site discrete Fourier transform and a Bogoliubov transformation. In particular, it is immediate to see that it conserves the canonical anti-commutation relations 
\be
\{\alpha_\mu(k),\alpha^\dagger_\nu(q)\}=\delta_{\mu,\nu}\delta_{k,q}\,.
\ee
Plugging \eqref{Eq:BogoTrans} in \eqref{Eq:Ham} we find 
\begin{align}
&H[J_2,\delta,U] = \sum_{\eta=\pm}\sum_{k > 0} \epsilon_\eta(k) \alpha^\dag_\eta(k) \alpha_\eta(k) \notag\\
&+ U \sum_{\bm\eta}\sum_{\bm k {>0}} V_{\bm \eta}({\bm k}) \alpha^\dag_{\eta_1}(k_1)\alpha^\dag_{\eta_2}(k_2)\alpha^{\phantom{\dag}}_{\eta_3}(k_3)\alpha^{\phantom{\dag}}_{\eta_4}(k_4)~,
\label{Eq:HFourierTransformed}
\end{align}
where bold symbols denote vectors with four components --- namely $\boldsymbol \eta=(\eta_1,\eta_2,\eta_3,\eta_4)$ and $\boldsymbol k=(k_1,k_2,k_3,k_4)$) --- we introduced the ``dispersion relation" 
\begin{multline}
\epsilon_\eta(k) = -2J_2\cos(2k) + 2\eta
\sqrt{\delta^2+(1-\delta^2)\cos^2(k)}\,,
\label{dispersion}
\end{multline}
and the ``vertex function" 
\begin{multline}
V_{\bm\eta}({\bm k}) = - \frac{1}{4} \sum_{P,Q\in S_2} {\rm sgn}(P) {\rm sgn}(Q) \\ \times V'_{\eta_{p_1}\eta_{q_1}\eta_{p_2}\eta_{q_2}}(k_{p_1},k_{q_1},k_{p_2},k_{q_2})\,,
\end{multline}
Here we denoted by $S_2$ the group of permutations of two elements, we introduced  
\begin{multline}
V'_{\bm \eta}(\bm k)\equiv \frac{e^{i(k_3-k_4)}}{2L}
\Big[(M_{12}+M_{34})\delta_{k_1-k_2+k_3-k_4,0}\\
+ (M_{12}-M_{34})\delta_{k_1-k_2+k_3-k_4\pm\pi,0} \Big]\,,
\label{Int}
\end{multline}
and finally
\be
M_{ab}\equiv\left(\eta_a\eta_be^{i \varphi_{k_a}(\delta)-i\varphi_{k_b}(\delta)}\right).
\ee
The physical interpretation of \eqref{Eq:HFourierTransformed} is transparent. It describes two species of fermions ($+$ and $-$) interacting via two-body scattering. 

\section{Setting}
\label{sec:prob}

In this work we are interested in the dynamics of the entanglement generated by a quantum quench comparing the case where the evolution is free, ${U=0}$, with that where it is weakly interacting, ${0<U\ll 1}$. As it is customary, we characterise the entanglement evolution computing the entanglement entropies of a finite subsystem $A$ in the thermodynamic limit. These are defined as 
\begin{equation}
S^{(\alpha)}_A(t)\equiv \frac{1}{1-\alpha}\log[{\rm tr}[\rho^\alpha_A(t)]]. 
\label{Sdef}
\end{equation}
Here $\rho_A(t)$ is the density matrix of the system at time $t$ reduced to the subsystem $A$; $\alpha$ is known as R\'enyi index, it is an arbitrary positive real number, although in many circumstances it is better to think of it as an integer.
In the limit $\alpha\to1$, the definition \eqref{Sdef} is the standard von Neumann entropy of $\rho_A$. 

Since we are interested in the dynamics of the entanglement \emph{generated} by the quench, it is convenient to prepare the system in a low entangled initial state. To this aim, we consider the standard quantum quench protocol~\cite{EF16}: we prepare the system in the \emph{ground state} $\ket{\Psi_0}$ of $H[J_{2i},\delta_i,U_i]$ and, at $t=0$, suddenly change the parameters
\be
(J_{20},\delta_0,U_0)\qquad\longmapsto\qquad (J_{2},\delta,U)\,.
\label{eq:quench}
\ee  
After the quench the state of the system is then given by 
\be
\ket{\Psi_t}=e^{i H[J_{2},\delta,U] t} \ket{\Psi_0},\qquad\qquad t>0\,.
\label{eq:timevolvingstate}
\ee
In this work we will always focus on the case $U_0=0$ to ensure that Wick's Theorem holds on the initial state (all connected cumulants with more than two fermionic operators vanish), as well as to have a well defined GGE for the integrable quench \cite{sc-14}. 
These are mandatory requirements for the applicability of our techniques. Moreover, for definiteness, we also set $J_{20}=0$.

Note that, due to the change in the dimerisation parameter $\delta$, the sudden quench \eqref{eq:quench}  produces non-trivial dynamics also for vanishing interactions and it is meaningful to ask how this is influenced by $U$. Another possibility to observe the same scenario preserving translational invariance is to break the particle number conservation in, e.g., the initial state. This, however, leads to a more complicated set of equations~\cite{BEGR:PRB}. 
In this sense the Hamiltonian \eqref{Eq:Ham} represents the minimal model to study prethermalization in weakly interacting systems.

Using the mapping \eqref{Eq:BogoTrans} we immediately find the following linear relations between the Bogoliubov fermions before --- $\{\alpha^\dag_{0,\eta}(k),\alpha_{0,\eta}(k)\}$ --- and after --- $\{\alpha^\dag_\eta(k),\alpha_\eta(k)\}$ --- the quench
\be
\alpha_{0,\pm}(k)=\frac{1\pm e^{i \Delta\varphi_k }}{2}\alpha_+(k)+\frac{1\mp e^{i \Delta\varphi_k }}{2} \alpha_-(k)\,,
\label{eq:prepost}
\ee
where 
\be
\Delta\varphi_k \equiv\varphi_k(\delta_0)-\varphi_k(\delta).
\ee 
A linear relation like \eqref{eq:prepost} leads to a simple representation of the ground state of the pre-quench Hamiltonian in terms of the post quench Bogoliubov fermions, see, e.g., Refs.~\cite{EF16, CEF1, SMT,sps-04}. Specifically, in our case we find
\be
\!\!\!\!\!\!\ket{\Psi_0} \!\!=\!\! \prod_{k>0} \!\!\left[{\cos\left(\!\frac{\Delta\varphi_k}{2}\!\right) \!\alpha^{\dag}_-(k)\!+\!i\sin\left(\!\frac{\Delta\varphi_k}{2}\!\right) \!\alpha^\dag_+(k)}\right]\!\! \ket{0}\!,
\label{eq:initialstate}
\ee
and $\ket{0}$ is the vacuum of the post-quench `Bogoliubov fermions $\{\alpha^{\dag}_\pm(k),\alpha^{\phantom{\dag}}_\pm(k)\}$.

\subsection{Free Quench}\label{subsec:integrablequench}

For $U=0$ the system is completely characterised by the occupation numbers, i.e. the expectation values of the (number conserving) fermion bilinears on the time evolving state~\eqref{eq:timevolvingstate}
\be
n_{\mu\nu}(k,t)\equiv \braket{\Psi_t|\alpha^\dag_{\mu}(k) \alpha^{\phantom{\dag}}_{\nu}(k)|\Psi_t}\,,\qquad \mu,\nu=\pm.
\label{eq:occnumb}
\ee
Since the Hamiltonian \eqref{Eq:HFourierTransformed} is quadratic we immediately have 
\be
n_{\mu\nu}(k,t)=n_{\mu\nu}(k) e^{i(\epsilon_{\mu}(k)-\epsilon_{\nu}(k))t}\,.
\label{eq:freeevolution}
\ee
Moreover using the form \eqref{eq:initialstate} of the initial state we find 
\be
\begin{split}
&n_{\pm\pm}(k) = \frac{(1\mp \cos \Delta\varphi_k )}{2},\\
&n_{\pm\mp}(k)= \frac{\mp i \sin(\Delta\varphi_k)}{2}\,.
\end{split}
\label{eq:occnumb0}
\ee
We see that the diagonal occupation numbers are conserved (independent of time) while the off-diagonal are oscillating. This means that the expectation values of local (in space) observables for large times --- and hence the Generalised Gibbs Ensemble --- are completely specified only by the former, the contribution of the latter vanishes in a power law fashion~\cite{EF16}.

\subsection{Weakly Non-Integrable Quench: Equations of Motion}\label{subsec:nonintegrablequench}

Whenever ${U\neq 0}$ the situation drastically complicates. Even though the occupation numbers completely characterise the system at $t=0$ (because of the initial state we chose), as soon as the time evolution begins, non-trivial correlations start to build up resulting in non-zero higher connected cumulants. This means that 
\be
\braket{\Psi_t|\alpha^\dag_{\mu_1}(k_1)\dots\alpha^\dag_{\mu_n}(k_n) \alpha^{\phantom{\dag}}_{\nu_1}(p_1) \dots \alpha^{\phantom{\dag}}_{\nu_n}(p_n)|\Psi_t}
\ee
cannot be expressed in terms of \eqref{eq:occnumb} anymore. Equivalently this means that the time evolving state ceases to be Gaussian. Moreover $n_{\mu\nu}(k,t)$ is no longer given by the simple expression \eqref{eq:freeevolution}: the evolution of the occupation numbers becomes non-trivially coupled to that of the higher cumulants. 

Here we shall assume that, when a Gaussian state is evolving according to an interacting Hamiltonian with $U\ll 1$, there exists a (parametrically large) time window over which higher cumulants \emph{remain small}. This leads to two key simplifications: (i) the state remains approximately Gaussian and  we continue to characterise it by means of \eqref{eq:occnumb}  (ii) the time evolution of the occupation numbers can be (approximately) determined. This assumption has been tested in Refs.~\cite{EsslerPRB14,  BEGR:PRL, BEGR:PRB} comparing the results for two- and four- point functions obtained in this way with tDMRG simulations and exact diagonalisation. Here we will further test it in the case of the entanglement entropy (proving its consistency with ED results).

To calculate the time evolution of \eqref{eq:occnumb} we use the equations of motion (EOM)~\cite{SK:EOM,INJPhys,NessiArxiv15,BEGR:PRL,ESY:QBE,LS:QBE, BEGR:PRL, BEGR:PRB, FMS:NIQBE, FMS:QBE, KBbook}. These are a set of evolution equations for $n_{\mu\nu}(k,t)$ obtained by truncating the infinite hierarchy of evolution equations of the connected cumulants. We refer to the literature, see e.g. Refs.~\cite{ESY:QBE, BEGR:PRB, FMS:NIQBE, KBbook}, for a detailed explanation of the method and only list the final equations. Specifically, we will follow the notation/conventions of Refs.~\cite{BEGR:PRL, BEGR:PRB} and consider two different truncation schemes respectively known as {\it First Order} and {\it Second Order}. Their names are motivated by the fact that --- at fixed $t$ and small $U$ --- these two schemes give results that are accurate respectively to the first and second order in $U$. This, however, does not mean that these two schemes are equivalent to a first or second order perturbative expansion. On the contrary, they correspond to the re-summation of a certain class of terms (infinitely many) in the perturbative series.

\subsubsection{First order EOM}

The First Order truncation scheme leads to the following equations~\cite{BEGR:PRB}
\begin{align}
&\!\!\!\!\!\!\partial_t {n}_{\mu\nu}(k,t) = i {\epsilon}_{\mu\nu}(k) n_{\mu\nu}(k,t)\notag\\ 
&\,\,\qquad+4iU \sum_{\{\gamma_i\}}\sum_{q{>0}} \big[V_{\gamma_1 \gamma_2 \gamma_3 \mu}(k,q,q,k) n_{\gamma_1\nu}(k,t)\! \notag\\ 
&\,\,\quad\qquad- V_{\nu \gamma_2 \gamma_3 \gamma_1}(k,q,q,k) n_{\mu\gamma_1}(k,t)\big] n_{\gamma_2\gamma_3}(q,t)\,.
\label{Eq:firstorderEOM1}
\end{align}
As shown in Refs.~\cite{BEGR:PRB, BrunoThesis} the results of these equations are equivalent those found in Ref.~\cite{EsslerPRB14} using the continuous unitary transformation (CUT) approach~\cite{W:CUT, UhrigCUT, UhrigCUT2,kehreinbook, MK:prethermalization}. In essence, these equations describe the approach to the prethermal regime and the relaxation to the non-thermal deformed GGE of Ref.~[\onlinecite{EsslerPRB14}]. 
More specifically, for $t\gg1$ one can expand the solution of \eqref{Eq:firstorderEOM1} as follows~\cite{BrunoThesis} 
\bw
\be
n_{\mu\nu}(k,t) = n_{\mu\nu}(k)e^{i (\epsilon^{\rm dr}_\mu(k)-\epsilon^{\rm dr}_\nu(k))t}+ U n_{\mu\nu}^{(1)}(k,t) e^{i (\epsilon^{\rm dr}_\mu(k)-\epsilon^{\rm dr}_\nu(k))t} + O(U^2),
\label{eq:perturbativesol}
\ee
where
\begin{align}
 \epsilon^{\rm dr}_\eta(k) =&  \,\epsilon_\eta(k) + 4 U \sum_\gamma \sum_{q>0} V_{\eta\gamma\gamma\eta}(k,q,q,k) n_{\gamma\gamma}(q)+O(U^2)\,,\label{eq:dE}\\	
n_{\mu\nu}^{(1)}(k,t) =&\, 4 \sum_{\{\gamma_i\}} \sum_{q>0} W_{\gamma_1\gamma_2\gamma_3\mu}(k,q) n_{\gamma_1\nu}(k) n_{\gamma_2\gamma_3}(q)\left[e^{i(\epsilon^{\rm dr}_{\gamma_1}(k)+\epsilon^{\rm dr}_{\gamma_2}(q)-\epsilon^{\rm dr}_{\gamma_3}(q)-\epsilon^{\rm dr}_{\mu}(k)) t}-1\right]\notag\\
&- 4 \sum_{\{\gamma_i\}} \sum_{q>0} W_{\nu\gamma_2\gamma_3\gamma_1}(k,q) n_{\mu \gamma_1}(k) n_{\gamma_2\gamma_3}(q)\left[e^{i(\epsilon^{\rm dr}_{\nu}(k)+\epsilon^{\rm dr}_{\gamma_2}(q)-\epsilon^{\rm dr}_{\gamma_3}(q)-\epsilon^{\rm dr}_{\gamma_1}(k)) t}-1\right],\\
W_{\gamma_1\gamma_2\gamma_3\gamma_4}(k_1,k_2)  = &  \lim_{B\to\infty} V_{\gamma_1\gamma_2\gamma_3\gamma_4}(k_1,k_2,k_2,k_1) \frac{1-e^{- B| \epsilon^{\rm dr}_{\gamma_1}(k_1)+\epsilon^{\rm dr}_{\gamma_2}(k_2)-\epsilon^{\rm dr}_{\gamma_3}(k_2)-\epsilon^{\rm dr}_{\gamma_4}(k_1)|}}{\epsilon^{\rm dr}_{\gamma_1}(k_1)+\epsilon^{\rm dr}_{\gamma_2}(k_2)-\epsilon^{\rm dr}_{\gamma_3}(k_2)-\epsilon^{\rm dr}_{\gamma_4}(k_1)},
\end{align}
\ew
and where also the higher orders in $U$ do not contain secular terms. This expansion can be obtained, e.g., by a perturbative solution of \eqref{Eq:firstorderEOM1} using the Method of Multiple Scales of Ref.~\cite{Shivamoggibook}.

Before moving to the Second order scheme we make two final remarks. First, the equation \eqref{Eq:firstorderEOM1} (and hence the solution~\eqref{eq:perturbativesol}) does not depend on $J_2$. Second, the solution \eqref{eq:perturbativesol} gives the following occupation numbers for the deformed GGE 
\bw
\be
n^{\rm dGGE}_{\mu\nu}(k)=\begin{cases}\displaystyle
n_{\mu\mu}(k) + 4 U \sum_{\gamma_1\gamma_2}\sum_{q>0} \left[W_{\mu\gamma_1\gamma_2\bar \mu}(k,q) n_{\mu\bar\mu}(k) - W_{\bar \mu\gamma_1\gamma_2\mu}(k,q) n_{\bar \mu\mu}(k)\right] n_{\gamma_1\gamma_2}(q) & \mu=\nu\\
\displaystyle 8 U \sum_{\gamma}\sum_{q>0} W_{\mu\gamma\gamma \bar \mu} (k,q) n_{\mu \bar\mu}(k)n_{\gamma\gamma}(q)  &\mu=\bar \nu
\end{cases}\,,
\label{eq:dGGE}
\ee
where $\bar\mu=-\mu$.
\subsubsection{Second Order EOM}

The Second Order truncation scheme (also known as second Born approximation~\cite{Bonitzbook}) leads to the following equations~\cite{BEGR:PRB} 
\begin{align}
\partial_t {n}_{\mu\nu}(k,t) =& 
i {\epsilon}_{\mu\nu}(k) n_{\mu\nu}(k,t)
+4iU\sum_{\{\gamma_i\}}\sum_{q{>0}} V_{\gamma_1 \gamma_2 \gamma_3 \mu}(k,q,q,k) 
e^{i \epsilon_{\gamma_1\nu}(k) t} e^{i \epsilon_{\gamma_2\gamma_3}(q) t} n_{\gamma_1\nu}(k) n_{\gamma_2\gamma_3}(q)~\nn
&-4iU\sum_{\{\gamma_i\}}\sum_{q{>0}} V_{\nu \gamma_2 \gamma_3 \gamma_1}(k,q,q,k) 
e^{i \epsilon_{\mu\gamma_1}(k) t} e^{i \epsilon_{\gamma_2\gamma_3}(q) t} n_{\mu\gamma_1}(k) n_{\gamma_2\gamma_3}(q)~\nn
& -U^2  \int_0^t \!\textrm{d} t'  \sum_{\vec{\gamma}}\sum_{k_1,k_2,k_3{>0}} \!\!\!\!\!L^{\vec{\gamma}}_{\mu\nu}(k_1,k_2,k_3; k; t-t') 
n_{\gamma_1\gamma_2}(k_1,t') n_{\gamma_3\gamma_4}(k_2,t')n_{\gamma_5\gamma_6}(k_3,t')\nn
& - U^2  \int_0^t \!\textrm{d}t'  \sum_{\bm \gamma}\sum_{k_1,k_2{>0}} \!\!\!\!K^{\bm\gamma}_{\mu\nu}(k_1,k_2; k; t-t') 
n_{\gamma_1\gamma_2}(k_1,t') n_{\gamma_3\gamma_4}(k_2,t')~.
\label{Eq:EOM}
\end{align}
where we denoted vectors of length six by $\vec
\gamma$, while the kernels are given by 
\begin{align}
&K^{\bm\gamma}_{\mu\nu}(k_1,k_2;k;t) \equiv 4 \sum_{k_3,k_4{>0}} \sum_{\eta,\eta'} 
X^{\gamma_1\gamma_3\eta\eta';\eta\eta'\gamma_4\gamma_2}_{{\bm k};{\bm k}'} (\mu,\nu;k;t),\\
&L^{\vec\gamma}_{\mu\nu}(k_1,k_2,k_3;k;t) \equiv 
8 \sum_{\eta}\sum_{k_4{>0}} X^{\gamma_1\gamma_3\gamma_6\eta;\eta\gamma_5\gamma_4\gamma_2}_{{\bm k};{\bm k}'}(\mu,\nu;k;t)
- 16\sum_{\eta} X^{\gamma_1\gamma_3\eta\gamma_4;\gamma_5\eta\gamma_6\gamma_2}_{k_1k_2k_1k_2;k_3k_1k_3k_1}(\mu,\nu;k;t).
\end{align}
\ew
Here $\bm k'$ is obtained from $\bm k$ by reversing the order of the elements and we introduced 
\begin{align}
&X^{{\bm\gamma};{\bm\eta}}_{{\bm k};{\bm q}}(\mu,\nu;q;t) \equiv
Y^{\bm\gamma}_{\mu\nu}({\bm k},q)V_{\bm\eta}({\bm q}) e^{i
  E_{\bm\gamma}({\bm k})t}  \nn  
&\qquad \qquad\qquad\qquad
  - ({\bm \gamma},{\bm k})\leftrightarrow({\bm
  \eta},{\bm q}),\\
&{E}_{\boldsymbol{\eta}}(\bm{q}) \equiv{\epsilon}_{\eta_1}(q_1)+{\epsilon}_{\eta_2}(q_2)-{\epsilon}_{\eta_3}(q_3)-{\epsilon}_{\eta_4}(q_4),\\
&{Y}_{\mu\nu}^{\boldsymbol{\eta}}(k,\bm{q}) \equiv \delta_{\nu,\eta_4}\delta_{k,q_4}{V}_{\eta_1\eta_2\eta_3\mu}(\bm{q}) +\delta_{\nu,\eta_3}\delta_{k,q_3}{V}_{\eta_1\eta_2\mu\eta_4}(\bm{q})\notag\\
&\quad-\delta_{\mu,\eta_2}\delta_{k,q_2}{V}_{\eta_1\nu\eta_3\eta_4}(\bm{q})-\delta_{\mu,\eta_1}\delta_{k,q_1}{V}_{\nu\eta_2\eta_3\eta_4}(\bm{q}).
\label{Eq:kernelsEOM}
\end{align}
As shown in Refs.~\cite{BEGR:PRL, BEGR:PRB}, for $0 \leq t \lesssim U^{-1}$ the two-point functions computed with \eqref{Eq:EOM} remain order $U$ close to those computed with \eqref{Eq:firstorderEOM1}, providing a perturbative correction. For ${t\gg U^{-1}}$, however, the solutions of \eqref{Eq:EOM} leave the prethermal plateau describing a drift of two point functions towards their thermal value~\cite{BEGR:PRL, BEGR:PRB}, although the EOM method is not guaranteed to capture all features emerging at asymptotically large times~\cite{KBCHH:15, LMMR:14}.

In the late-time regime the time integrals in \eqref{Eq:EOM} can be simplified~\cite{BEGR:PRB} obtaining a local-in-time Quantum Boltzmann Equation (QBE)~\cite{FMS:NIQBE, FMS:QBE, ESY:QBE,LS:QBE} (see also~\cite{FGV:19, DBD:20, LWGV:20} for recent generalizations of the QBE to treat interacting integrable systems). More precisely considering the scaling limit 
\be
 U\to0\quad\text{and}\quad t\to\infty\quad \text{with}\quad \tau = t
 U^2 \quad\text{fixed}\,,
 \label{Eq:Boltzscallim}
\ee
and filtering out highly oscillating terms (which do not contribute to local observables) one obtains the following equation for the diagonal occupation numbers~\cite{BEGR:PRB} 
\bw
\be
\partial_{\tau}{n}_{\mu\mu}(k,\tau) =  - \sum_{\eta,\gamma}\sum_{p,q{>0}} 
\widetilde K^{\gamma\eta}_{\mu}(p,q|k) n_{\gamma\gamma}(p,\tau) n_{\eta\eta}(q,\tau)-\sum_{\gamma,\eta,\varphi}\sum_{p,q,r{>0}} \
\!\!\!\!\widetilde L^{\gamma\eta\varphi}_{\mu}(p,q,r|k)n_{\gamma\gamma}(p,\tau)
n_{\eta\eta}(q,\tau)n_{\varphi\varphi}(r,\tau)\,.
\label{Eq:BB}
\ee
Here the kernels are given by 
\be
\begin{split}
\widetilde K^{\gamma_1\gamma_2}_{\alpha}(k_1,k_2|q) &\equiv 4 \sum_{k_3,k_4{>0}} \sum_{\nu,\nu'}  
\widetilde X^{\gamma_1\gamma_2\nu\nu'|\nu\nu'\gamma_2\gamma_1}_{{\bm k}|{\bm k}'} (\alpha|q),\\
\widetilde L^{\gamma_1\gamma_2\gamma_3}_{\alpha}(k_1,k_2,k_3|q) & \equiv 8
\sum_{\nu}\sum_{k_4{>0}} \widetilde X^{\gamma_1\gamma_2\gamma_3\nu|\nu\gamma_3\gamma_2\gamma_1}_{{\bm k}|{\bm k}'}(\alpha|q)- 16\sum_{\nu} \widetilde X^{\gamma_1\gamma_2\nu\gamma_2|\gamma_3\nu\gamma_3\gamma_1}_{k_1k_2k_1k_2|k_3k_1k_3k_1}(\alpha|q),\\
\widetilde X^{{\bm\gamma}|{\bm\alpha}}_{{\bm k}|{\bm q}}(\alpha|q) &\equiv
Y^{\bm\gamma}_{\alpha\alpha}({\bm k},q)V_{\bm\alpha}({\bm q}) D(
E_{\bm\gamma}({\bm k}))- ({\bm \gamma},{\bm k})\leftrightarrow({\bm \alpha},{\bm q})\,,\\
D(E) &\equiv\lim_{\xi\to0}\frac{i}{E+i \xi}\,.
\end{split}
\label{Eq:Boltkernels}
\ee
\ew
As initial value for the Quantum Boltzmann Equation one takes the diagonal occupation numbers $\{n_{\pm\pm}(k,t_0)\}$ produced by the second order EOM \eqref{Eq:EOM} for large enough $t_0$ (corresponding to the deformed GGE values). 

It can be verified (see, e.g., \cite{FMS:NIQBE, BEGR:PRB}) that non-interacting Fermi-Dirac distribution (with arbitrary $\beta$ and $\mu$) is always a stationary solution of the QBE. Moreover, for non-integrable models the latter is believed to be the only stationary solution~\cite{FMS:NIQBE}. The specific values of temperature and chemical potential can be determined from the initial conditions by noting that the QBE conserves number density and the kinetic energy. This means that in our case we expect the QBE to describe relaxation to a ``free Gibbs Ensemble'', reproducing  the solution of the second order EOM \eqref{Eq:EOM} only up to $O(U)$ corrections. Importantly this is enough to observe the transition between GGE and thermal values because their difference is $O(U^0)$, i.e., they are different even for $U=0$. 

\section{Entanglement Entropy from EOM}
\label{sec:EE}

We compute the entropies under the assumption that the state remains approximately Gaussian for $t>0$. This assumption is in principle stronger than the one used to derive \eqref{Eq:EOM}. Indeed, in the derivation of \eqref{Eq:EOM} four-particle cumulant is approximated by a non-zero value~\cite{BEGR:PRB}. We expect that, however, in the time window of validity of the EOM this approximation gives the leading order (in $U$) of the entropies.    

Under the Gaussian-state assumption we can directly compute the entanglement entropy from the correlation matrix
\be
\begin{split}
C_{ij}(t)&= \braket{\Psi_t|c^\dag_i c^{\phantom{\dag}}_j|\Psi_t} \\
&= \frac{1}{L} \sum_{k>0} \sum_{\mu,\nu = \pm} 
\gamma^*_\mu(k,i) \gamma^{\phantom{\dag}}_\nu(k,j) n_{\mu\nu}(k,t)\,,
\end{split}
\label{Eq:GreenFun}
\ee
as follows~\cite{lrv-03, peschel2003,pe-09}
\be
S^{(\alpha)}_A(t)\equiv \frac{1}{1-\alpha}{\rm tr}\log[C^\alpha+(1-C)^\alpha].
\ee 
Note that this form is particularly simple because the correlations $\braket{\Psi_t|c^\dag_i c^{\dag}_j|\Psi_t}$ and $\braket{\Psi_t|c_i c_j|\Psi_t}$ are zero at all times. This is a consequence of both the initial state and the Hamiltonian being $U(1)$-invariant (the initial state has a fixed number of fermions and the Hamiltonian conserves it).

To test the EOM predictions, we compute the entropy of the thermal state using exact diagonalisation (ED). More precisely, we evaluate numerically 
\be
S^{(\alpha)}_{\rm th} = \frac{1}{1-\alpha}\log{\rm tr}[\rho_{\rm th}^\alpha]\,,
\label{eq:thermalS}
\ee
in two different ways to estimate the finite-size corrections. First we used the canonical representation for the thermal density matrix 
\be
\rho_{\rm th}=\frac{e^{-\beta H[J_2,\delta,U]}}{{\rm tr}[e^{-\beta H[J_2,\delta,U]}]}\,,
\ee
where the temperature is fixed by requiring 
\be
{\rm tr}[\rho_{\rm th} H[J_2,\delta,U]]= \braket{\Psi_0|H[J_2,\delta,U]|\Psi_0}\,.
\ee 
In this representation the trace in \eqref{eq:thermalS} must be reduced to states with fixed particle number. Namely to a basis of the eigenspace of 
\be
N=\sum_{l=1}^L c^\dag_l c^{\phantom\dagger}_{l}\,,
\ee
corresponding to the eigenvalue $L/2$. Next, we consider the grand-canonical representation of $\rho_{\rm th}$, namely 
\be
\rho_{\rm th}=\frac{e^{-\beta (H[J_2,\delta,U]-\mu N)}}{{\rm tr}[e^{-\beta (H[J_2,\delta,U]-\mu N)}]}\,,
\ee
where temperature and chemical potential are fixed by requiring 
\be
\begin{split}
{\rm tr}[\rho_{\rm th} H[J_2,\delta,U]] &= \braket{\Psi_0|H[J_2,\delta,U]|\Psi_0},\\
{\rm tr}[\rho_{\rm th} N] &= \braket{\Psi_0|N|\Psi_0}=L/2\,. 
\end{split}
\ee
Finally, we extrapolate the results for $L\to\infty$ assuming 
\be
S^{(\alpha)}_{\rm th}(L) = S^{(\alpha)}_{\rm th}(\infty)  + \frac{A^{(\alpha)}}{L}  + O\left(\frac{1}{L^2}\right)\,,
\ee
where $A^{(\alpha)}$ is a constant determined through a linear fit. This procedure is sometimes plagued by even-odd effects in the ED data, limiting its range of applicability.

\begin{figure*}[t]
\begin{tabular}{l}
\includegraphics[width=0.425\textwidth]{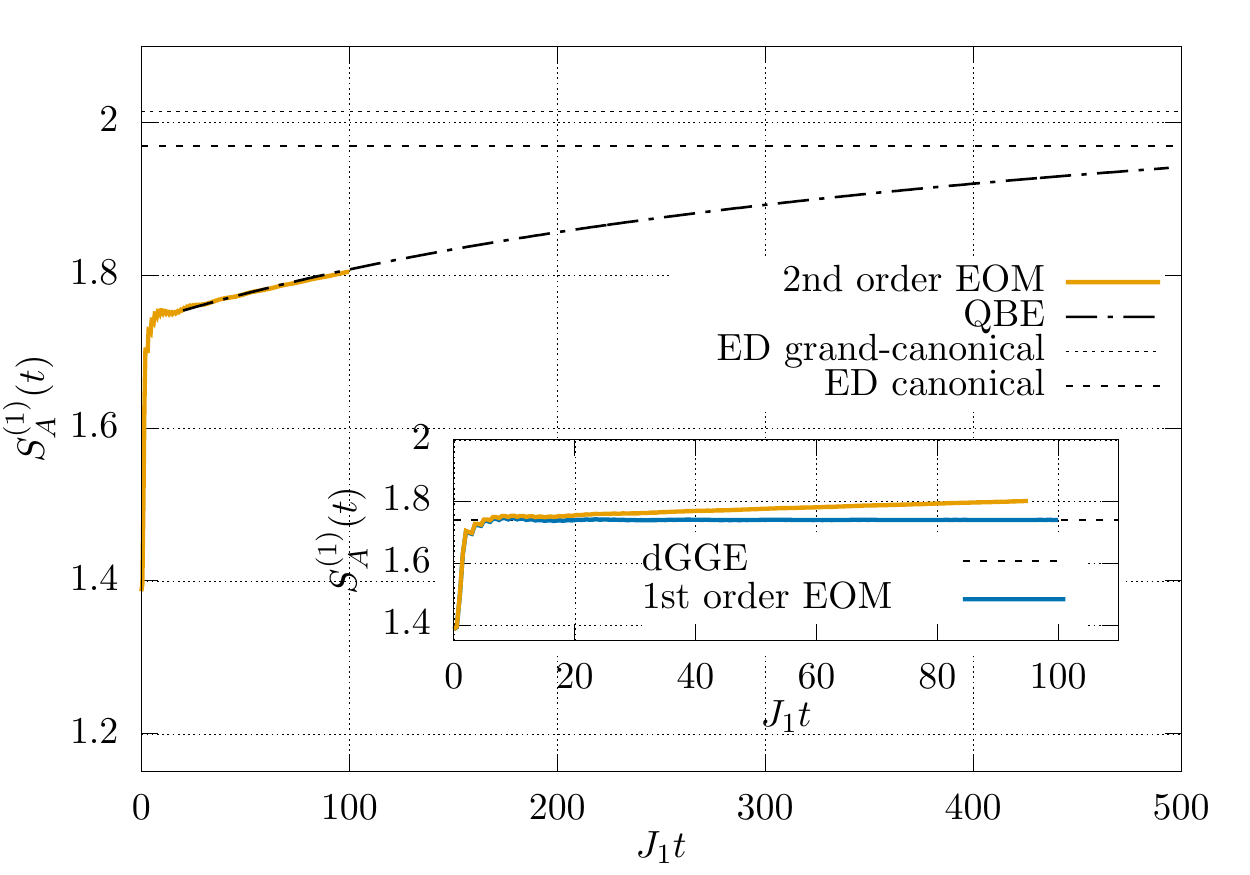} \qquad 
\includegraphics[width=0.425\textwidth]{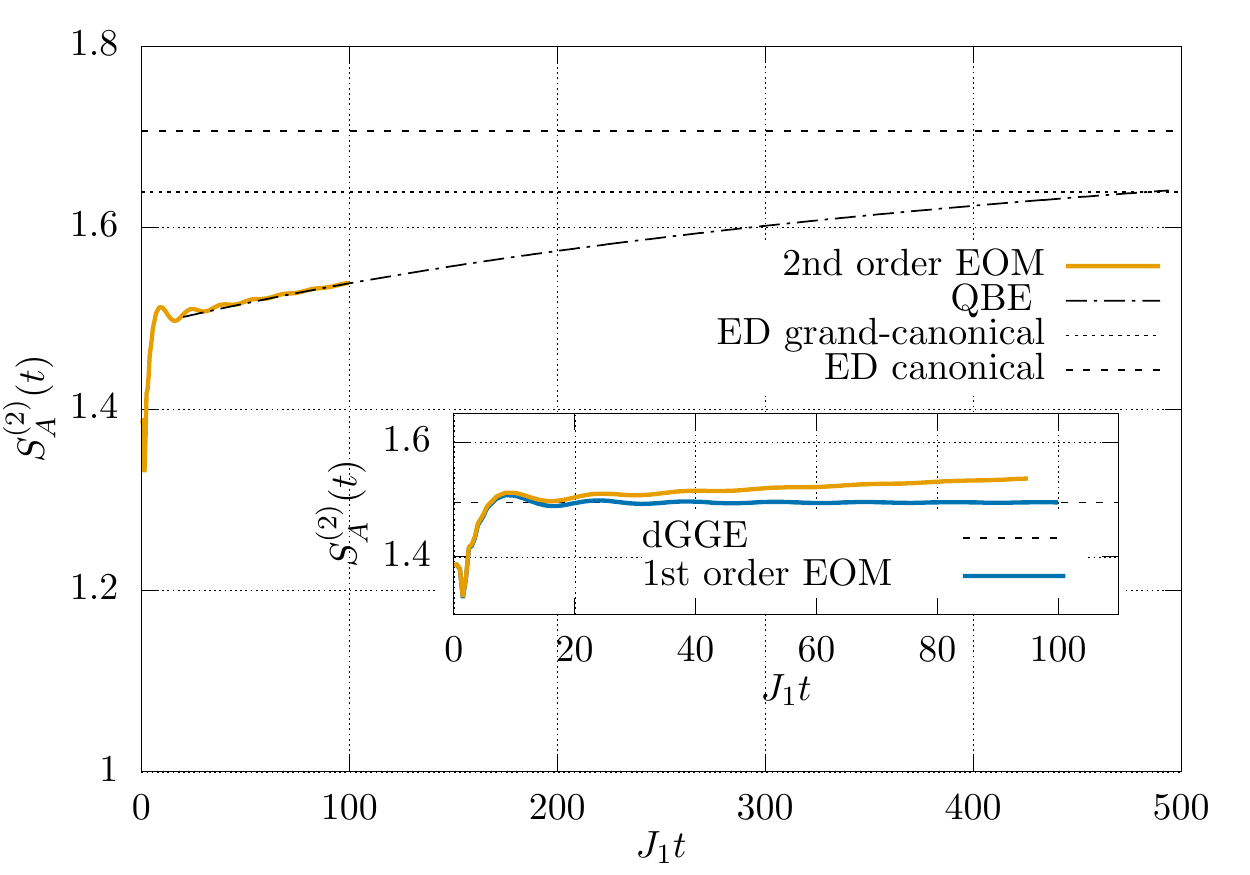} 
\end{tabular}
\caption{Time-evolution of the von Neumann entropy (left) and Renyi-2 entropy (right) after the quench \eqref{eq:quench} with parameters $J_{20}=0$, $\delta_0=0.8$, $U_0=0$, $J_{2}=0.65$, $\delta=0.1$, $U=0.1$, and $|A|=4$. Different lines report the predictions of the different methods and the insets present the late time behaviour computed via Quantum Boltzmann Equation (initialised at time $t_0=20.1$). The difference between canonical and grand-canonical prediction gives an estimation of the error in the extrapolation of the ED results.} 
\label{Fig:EntropiesSmallSystems2}
\end{figure*}

\begin{figure*}[t]
\begin{tabular}{l}
\includegraphics[width=0.425\textwidth]{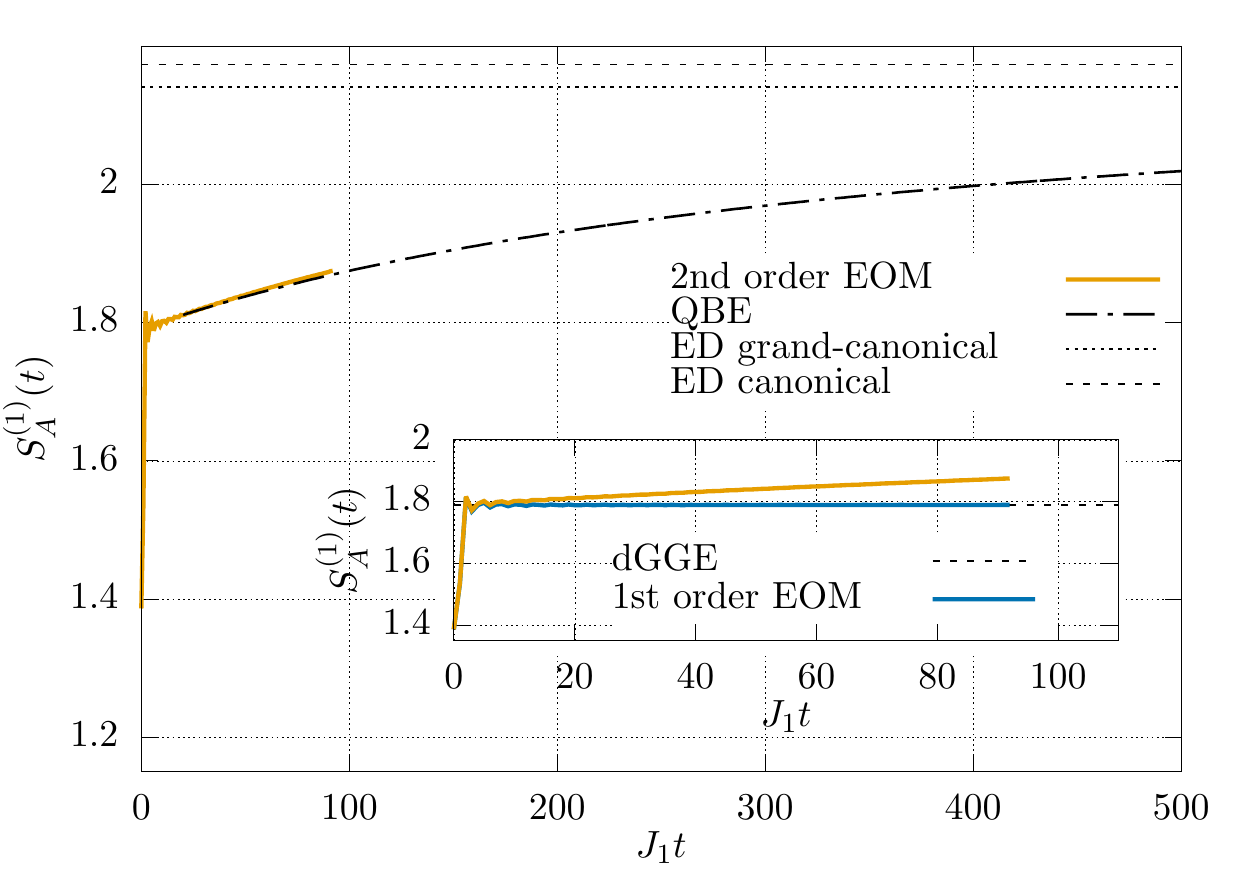} \qquad 
\includegraphics[width=0.425\textwidth]{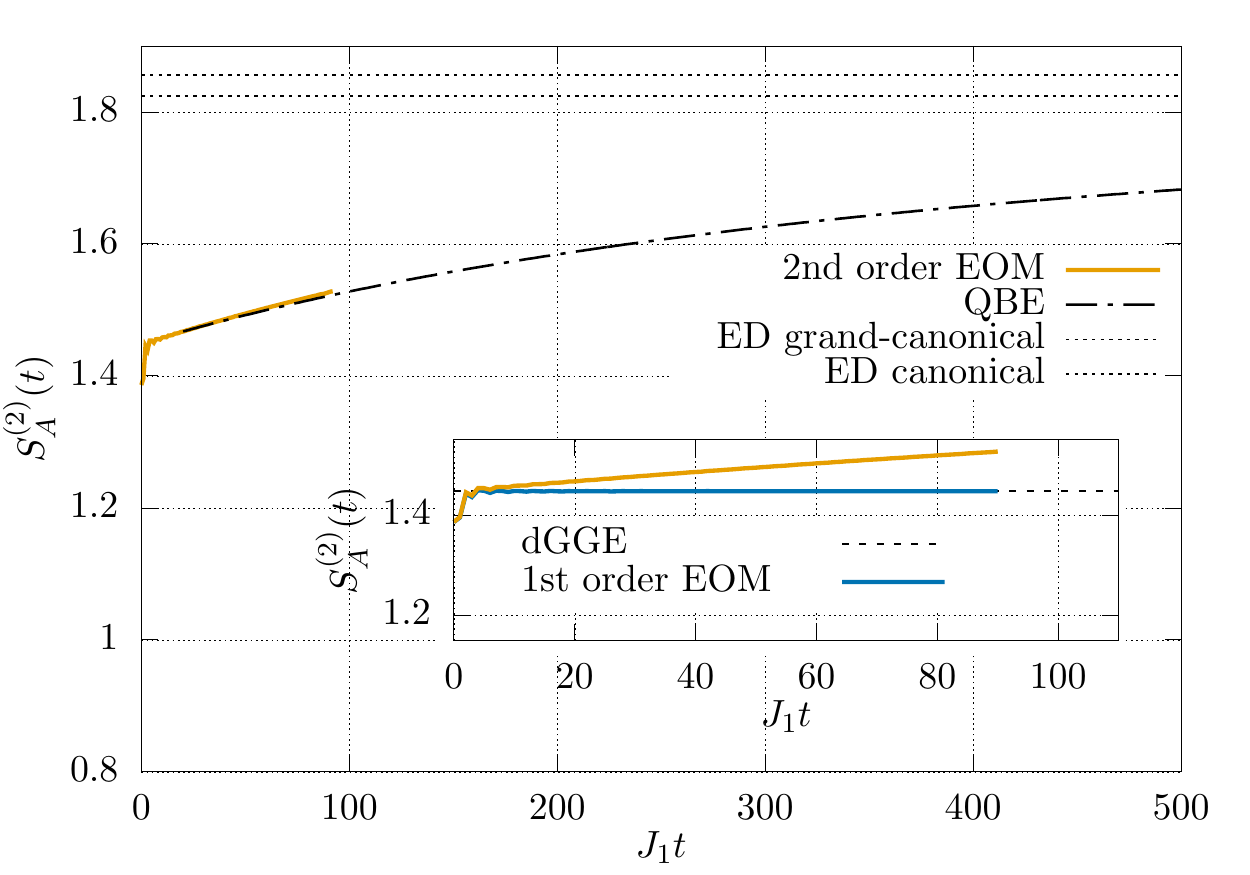} 
\end{tabular}
\caption{Time-evolution of the von Neumann entropy (left) and Renyi-2 entropy (right) after the quench \eqref{eq:quench} with parameters $J_{20}=0$, $\delta_0=0.8$, $U_0=0$, $J_{2}=0.65$, $\delta=0$, $U=0.1$, and $|A|=4$. Different lines report the predictions of the different methods and the insets present the late time behaviour computed via Quantum Boltzmann Equation (initialised at time $t_0=20.1$). The difference between canonical and grand-canonical prediction gives an estimation of the error in the extrapolation of the ED results.} 
\label{Fig:EntropiesSmallSystems3}
\end{figure*}

\begin{figure*}[t]
\begin{tabular}{l}
\includegraphics[width=0.425\textwidth]{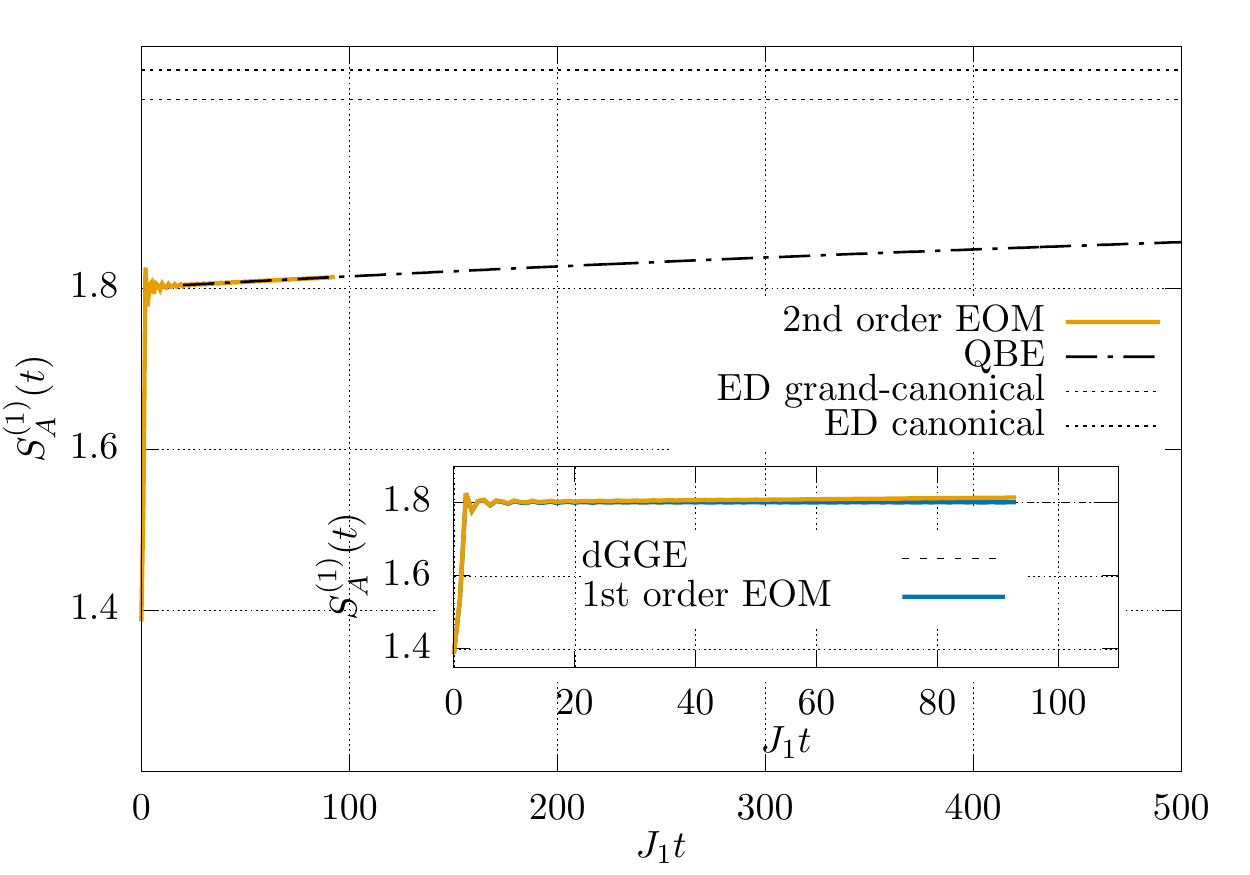} \qquad 
\includegraphics[width=0.425\textwidth]{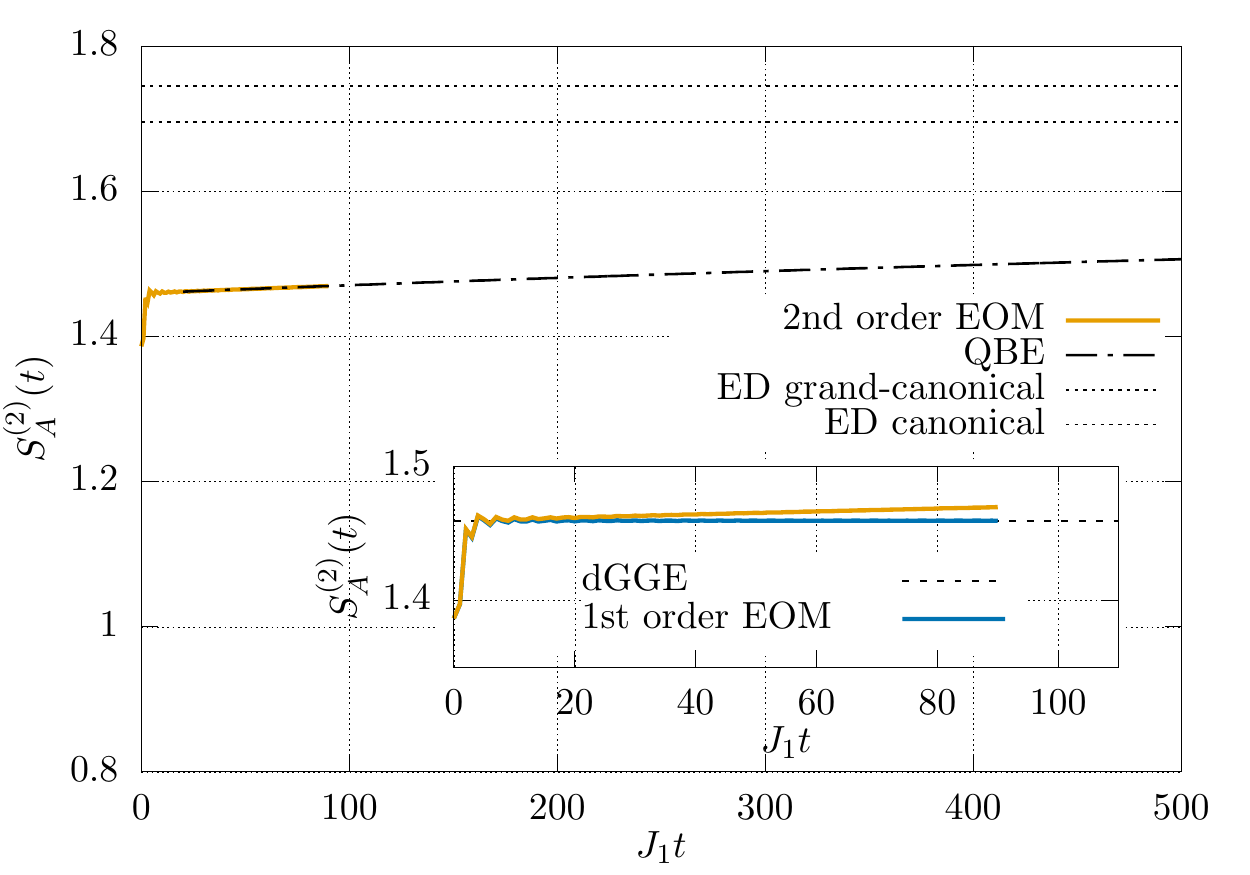} 
\end{tabular}
\caption{Time-evolution of the von Neumann entropy (left) and Renyi-2 entropy (right) after the quench \eqref{eq:quench} with parameters $J_{20}=0$, $\delta_0=0.8$, $U_0=0$, $J_{2}=0.5$, $\delta=0$, $U=0.05$, and $|A|=4$. Different lines report the predictions of the different methods and the insets present the late time behaviour computed via Quantum Boltzmann Equation (initialised at time $t_0=20.1$). The difference between canonical and grand-canonical prediction gives an estimation of the error in the extrapolation of the ED results.} 
\label{Fig:EntropiesSmallSystems1}
\end{figure*}

Some representative examples of our results are reported in Figs.~\ref{Fig:EntropiesSmallSystems2}\,--\,\ref{Fig:EntropiesSmallSystems1}. We see that, while the first order EOM~\eqref{Eq:firstorderEOM1} predict relaxation to a deformed GGE, the solution of the second order EOM \eqref{Eq:EOM} shows a slow drift towards the thermal value computed by ED. Such a drift occurs on time-scales $t\sim U^{-2}$ and is well described by the QBE~\eqref{Eq:BB}, in agreement with the findings of Refs.~\cite{BEGR:PRL, BEGR:PRB}. In particular, this implies that for large enough times the drift can be described by an exponential (as it is customary the characteristic time can be obtained by linearising the QBE~\cite{SK:EOM}). 
 
Finally, we see that the speed of the drift is highly influenced by the value of the next-neighbour hopping term $J_2$. This has been explained in~\cite{BEGR:PRL, BEGR:PRB} (see also Refs.~\cite{FMS:NIQBE, BK:rates}) by noting that $J_2$ opens more {\it scattering channels} for the quasiparticles. To understand this point it is convenient to look at the Boltzmann equation~\eqref{Eq:BB}. We see that the only scattering processes contributing to the r.h.s. of this equation are those conserving energy and momentum (modulo $\pi$) --- the energy conservation is enforced by $D(E)$ while the momentum conservation by the vertex $V_{\bm\alpha}({\bm q})$. A non-zero $J_2$ allows for more inelastic solutions of these constraints, namely for processes
\be
\begin{tikzpicture}[baseline=(current  bounding  box.center), scale=1.5]
\draw[thick] (-4.25,0.5) -- (-3.25,-0.5);
\draw[ thick] (-4.25,-0.5) -- (-3.25,0.5);
\draw[->, thick] (-4.25, -0.5) -- (-4,-0.25);
\draw[->, thick] (-3.25, -0.5) -- (-3.5,-0.25);
\draw[->, thick] (-3.75, 0) -- (-3.4,0.25+0.1);
\draw[->, thick] (-3.75, 0) -- (-4.1,0.25+0.1);
\draw[->, thick] (-3.25, -0.5) -- (-3.5,-0.25);
\draw[ thick, fill=gray, rounded corners=2pt] (-3.75,0) circle (0.25cm);
\Text[x=-4.25,y=-0.75]{}
\Text[x=-4.5,y=-0.75]{$(\alpha_i, k_i)$}
\Text[x=-3,y=-0.75]{$(\beta_i, p_i)$}
\Text[x=-4.5,y=0.75]{$(\alpha_f, k_f)$}
\Text[x=-3,y=0.75]{$(\beta_f, p_f)$}
\draw[|->, thick] (-5.75, -0.5) -- (-5.75,0.5);
\Text[x=-5.95,y=0.3]{$t$}
\end{tikzpicture},
\ee
with $\{(\alpha_i, k_i), (\beta_i, p_i) \}\neq\{ (\alpha_f, k_f),(\beta_f, p_f)\}$. These are the scattering processes able to modify the momentum distribution.

\section{EOM and Quasi-particle picture}
\label{sec:quasiparticles}

In the integrable case the initial state \eqref{eq:initialstate} can be viewed as a source of $+$ and $-$ quasiparticles, where those with the same momentum $k$ are correlated. This can be understood by noting that the eigenstates of $H[J_2,\delta,0]$ with non-zero overlap with $\ket{\Psi_0}$ feature only one of the modes $(+, k)$ and $(-, k)$, such a non-trivial microscopic constraint generates the correlation~\cite{triplets}. For $t,|A|\gg1$ we can then describe the evolution of the entanglement entropies at the leading order using the quasiparticle picture~\cite{CA, CC}. Writing the prediction in the case of correlated pairs with different velocities we have  
\begin{equation}
\!\!\!\!S^{(\alpha)}_{{\rm q}, A}(t) \!=\!\! \int_{0}^\pi \!\!\!\!\!{\rm d}k\,
{\rm min}[|v_+(k)-v_-(k)| t, |A|] s_{\alpha}[n_{++}(k)],
\label{eq:quasiparticles}
\end{equation}
where $n_{++}(k)$ is the conserved (for $U=0$) diagonal occupation number of $+$ particles (cf.~\eqref{eq:occnumb0}), the group velocities are given by 
\be
\begin{split}
v_\pm(k) \equiv \partial_k\epsilon_\pm(k)= &\frac{\mp (1-\delta^2)\sin(2k)}{\sqrt{\delta^2+(1-\delta^2)\cos^2(k)}} \\
&+ 4 J_2 \sin(2 k),
\end{split}
\ee
and we introduced  
\begin{align}
s_{\alpha}[x] = \frac{x^\alpha+(1-x)^{\alpha}}{\pi(1-\alpha)}\,.
\label{eq:yangyang}
\end{align}
Note that since 
\be
\label{eq:symmetryPsi0}
n_{++}(k)+n_{--}(k)= 1\,,
\ee
we have 
\be
s_{\alpha}[n_{++}(k)]=s_{\alpha}[n_{--}(k)]\,,
\label{eq:phsimmetry}
\ee
and the prediction is symmetric in $+$ and $-$. Finally, since \eqref{eq:quasiparticles} only depends on the difference of $v_-(k)$ and $v_+(k)$, it is independent of $J_2$. 

\begin{figure}[b]
\begin{tabular}{l}
\includegraphics[width=0.5\textwidth]{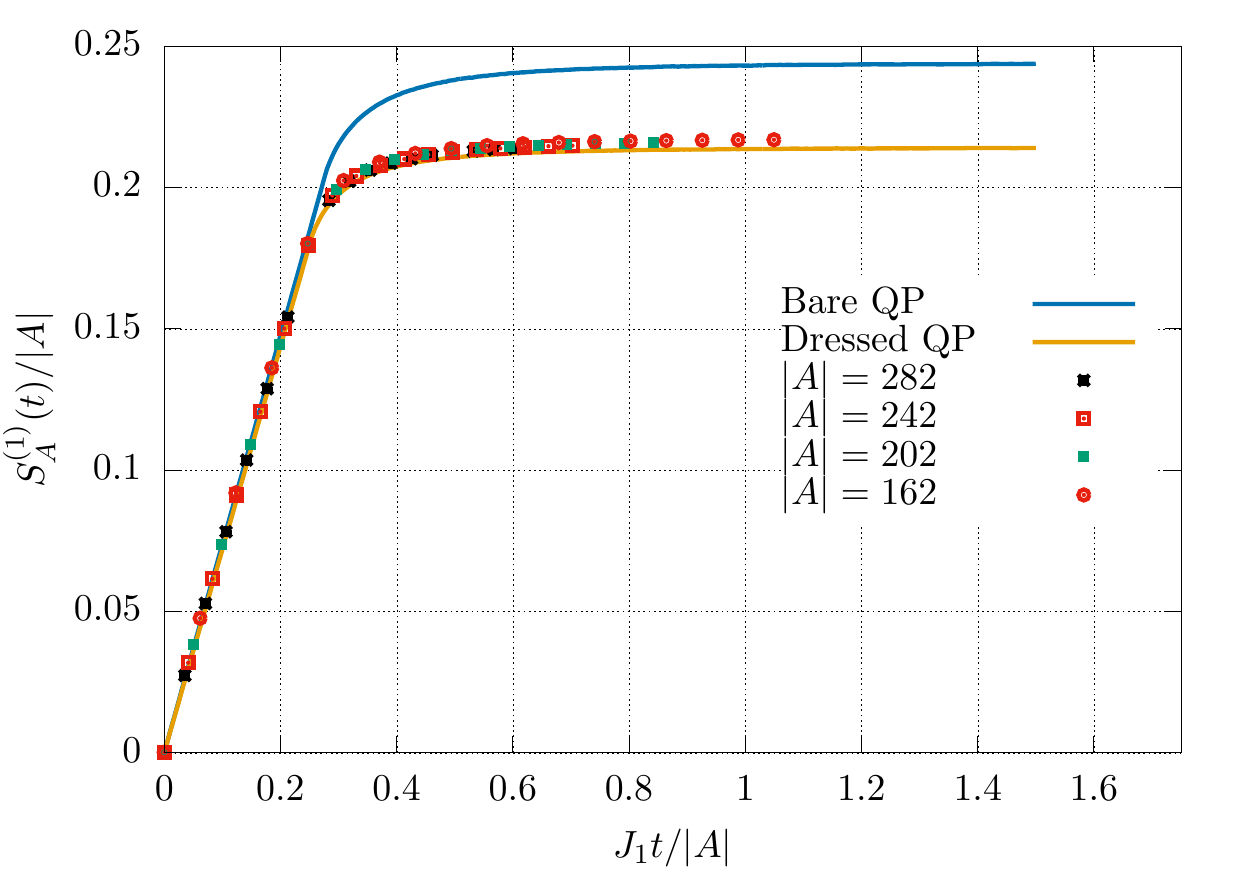} 
\end{tabular}
\caption{Comparison between the solution of the first order EOM \eqref{Eq:firstorderEOM1} and the prediction of the quasiparticle picture for the rescaled entanglement entropy $S_A^{(1)}$ (von Neumann) after a quench \eqref{eq:quench} with parameters $J_{20}=0$, $\delta_0=0.8$, $U_0=0$, $J_{2}=0$, $\delta=0.1$, $U=0.4$. 
The blue solid line is the prediction of bare quasiparticles while the orange solid line is the one of dressed quasiparticles. Points are the (rescaled) von Neumann entropies for $|A|=162,202,242,282$ obtained through Eq.~\eqref{Eq:firstorderEOM1}.} 
\label{Fig:DressedQP}
\end{figure}

 \begin{figure*}[t]
\includegraphics[width=0.45\textwidth]{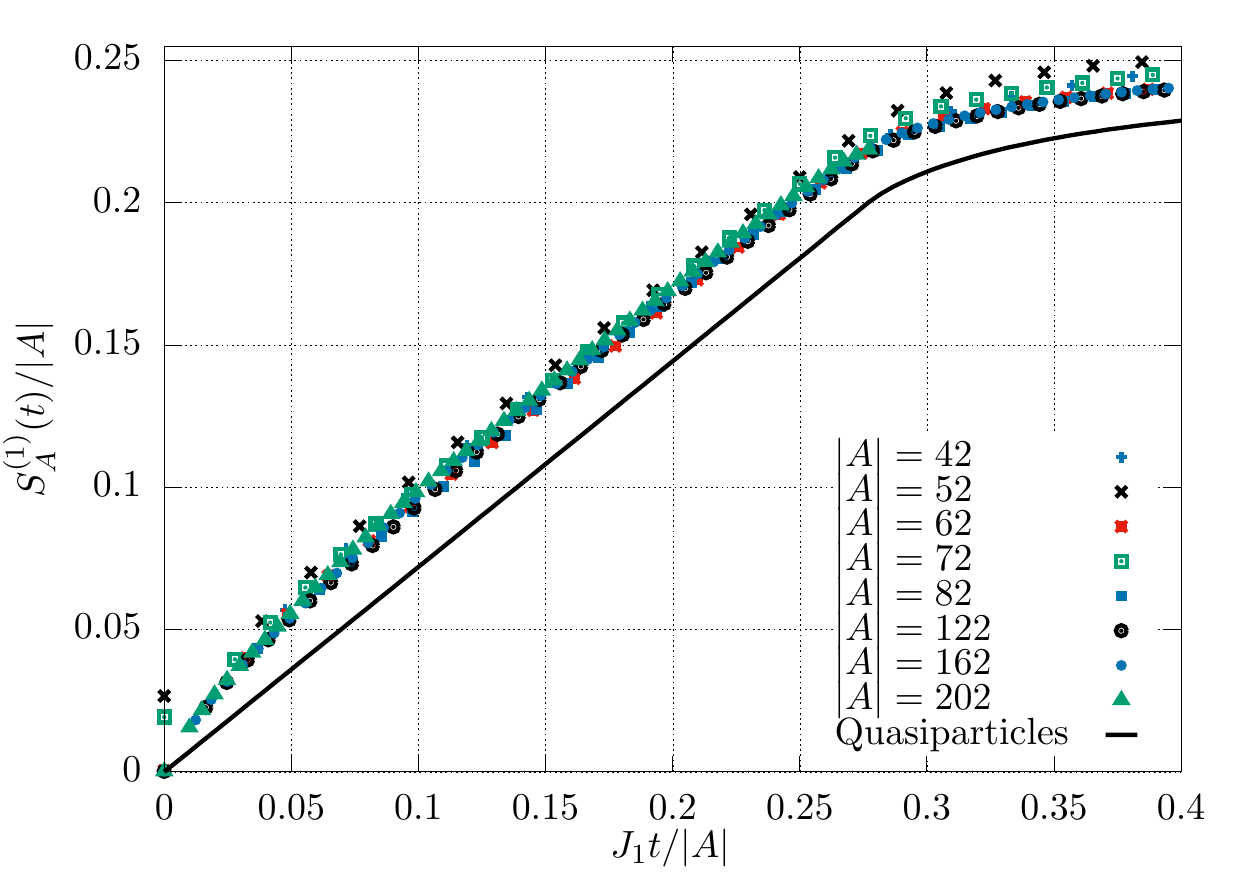} \includegraphics[width=0.45\textwidth]{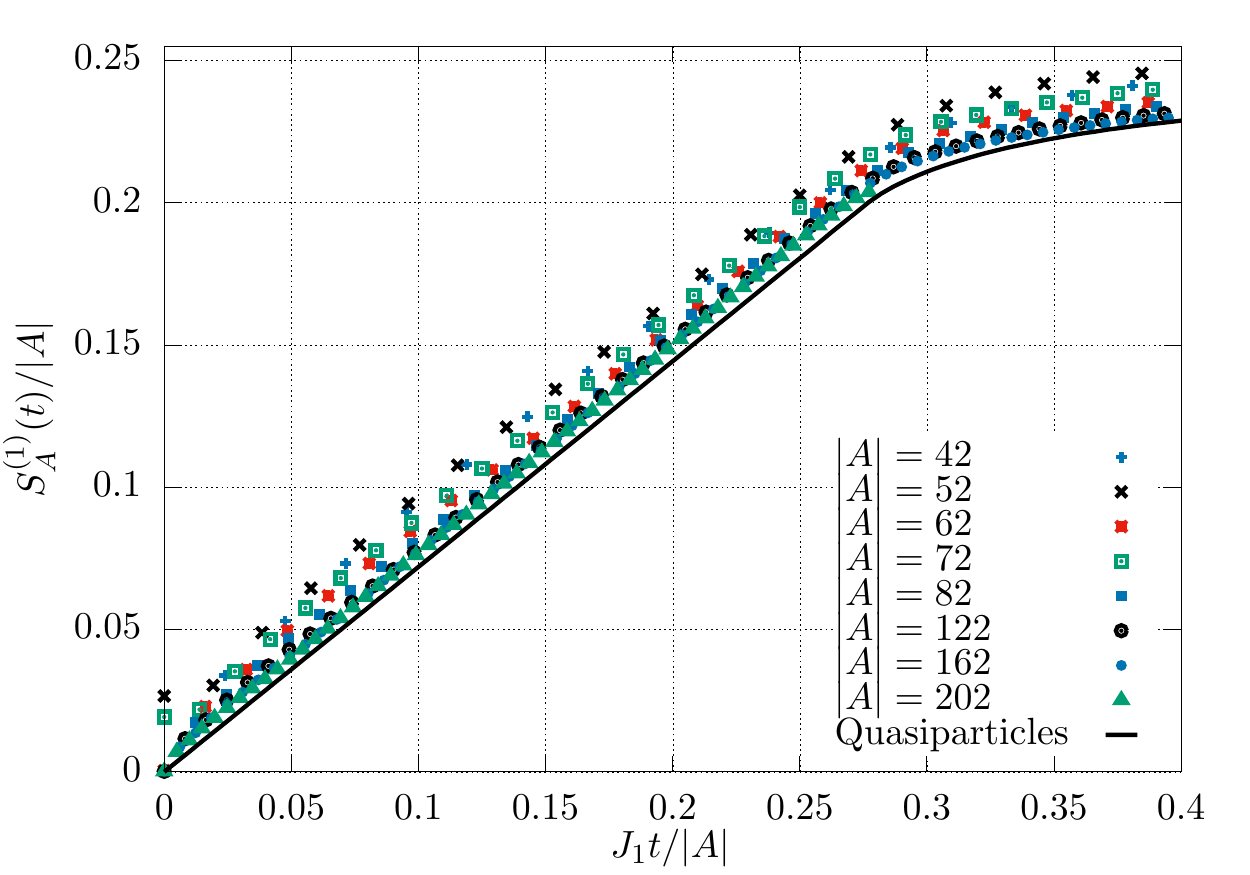} 
\caption{Comparison between the scaling of the solutions of the second order EOM \eqref{Eq:EOM} (left) and that of the first order EOM~\eqref{Eq:firstorderEOM1} (right) where the black line is the ``scattering quasiparticle" prediction of Eq.~\eqref{eq:quasiparticlesBolt}. The parameters of the quench are fixed to $J_{20}=0$, $\delta_0=0.8$, $U_0=0$, $J_{2}=0.5$, $\delta=0.1$, $U=0.1$. The collapse of the first order data is attained for $\ell_{\rm min}\approx 200$, while $\bar v U^{-2}\approx 100$.} 
\label{Fig:DifferentScaling}
\end{figure*}

The question that we want to address here is whether a similar quasiparticle prediction can be devised for $0<U \ll 1$. In our simple framework everything is fully specified by $n_{\mu\nu}(k, t)$, therefore to understand whether a quasiparticle picture can work we need to look at how these quantities depend on time. Inspecting the perturbative solution \eqref{eq:perturbativesol} of the first order EOM~\eqref{Eq:firstorderEOM1} we see that at large enough times it has the same form as the free one (cf. \eqref{eq:freeevolution}) with the replacements
\be
\epsilon_\eta(k) \mapsto \epsilon^{\rm dr}_\eta(k),\qquad n_{\mu\nu}(k) \mapsto n^{\rm dGGE}_{\mu\nu}(k),
\ee
where $\epsilon^{\rm dr}_\eta(k)$, $n^{\rm dGGE}_{\mu\nu}(k)$ are defined respectively in Eqs.~\eqref{eq:dE} and \eqref{eq:dGGE} and we neglected $O(U^2)$ corrections. This means that for ${0\leq t\lesssim U^{-1}}$, i.e. in the time regime where the first order EOM give a good description of the results, we can define a quasiparticle picture by replacing  $v_\pm(k)$ and $n_{\pm\pm}(k)$ with their dressed counterparts obtained from Eqs.~\eqref{eq:dE} and \eqref{eq:dGGE}. Namely 
\begin{equation}
\!\!\!\!S^{(\alpha)}_{{\rm dq}, A}(t) \!=\!\! \int_{0}^\pi \!\!\!\!\!{\rm d} k\, {\rm min}[|v^{\rm dr}_+(k)\!-\!v^{\rm dr}_-(k)| t, |A|] s_{\alpha}[n^{\rm dr}_{++}(k)],
\label{eq:dquasiparticles}
\end{equation}
with 
\begin{align}
 v^{\rm dr}_\eta(k)  =\partial_k \epsilon^{\rm dr}_\eta(k),\qquad n_{\mu\mu}^{\rm dr}(k) =n^{\rm dGGE}_{\mu\nu}(k). \label{eq:dressedn} 
\end{align}
A comparison between bare and dressed quasiparticle predictions and the solution of the first order EOM~\eqref{Eq:firstorderEOM1} is reported in Fig.~\ref{Fig:DressedQP}. Note that in this case $n^{\rm dGGE}_{+-}(k)\neq 0$ and one would need to ``diagonalise" the quasiparticle occupations as in Ref.~\cite{intertwined}. This effect, however, gives a sub-leading correction in $U$ and since we are working at $O(U)$ we can neglect it. Finally, we remark that a quasiparticle picture valid for times ${t\sim U^{-1}}$ can be devised also when the system displays pre-relaxation, see Ref.~\cite{FC15}. Namely, when, due to some special symmetries of the unperturbed Hamiltonian, the deformed GGE becomes (slowly) time-dependent for $0\leq t \lesssim U^{-1}$~\cite{BF15}.

For $t\gg U^{-1}$ the integrability breaking begins to dominate and, accordingly, the quasiparticles start to scatter inelastically approaching the equilibrium state. Consequently, the occupation numbers $n_{\pm\pm}(k)$ evolve from their deformed GGE values $n^{\rm dGGE}_{\pm\pm}(k)$ to their thermal values. In this situation it is natural to distinguish two different regimes for the behaviour of the entanglement of the subsystem~$A$
\be
\text{(i)}\,\, |A|\ll \bar v U^{-2}\qquad \text{and} \qquad \text{(ii)}\,\, |A|\gg  \bar v U^{-2},
\ee
where $\bar v$ is the minimal velocity of the quasiparticles giving a relevant contribution to the entanglement. In other words $\bar v$ is the maximal $v$ such that $S_{{\rm q}, A}^{(\alpha)}(|A|/v)$ essentially equals its saturation value. For the quenches considered here $\bar v\approx 1$ (for example $\bar v\approx 1.2$ in the case of Fig.~\ref{Fig:DressedQP}).

\begin{figure*}[t]
\includegraphics[width=0.45\textwidth]{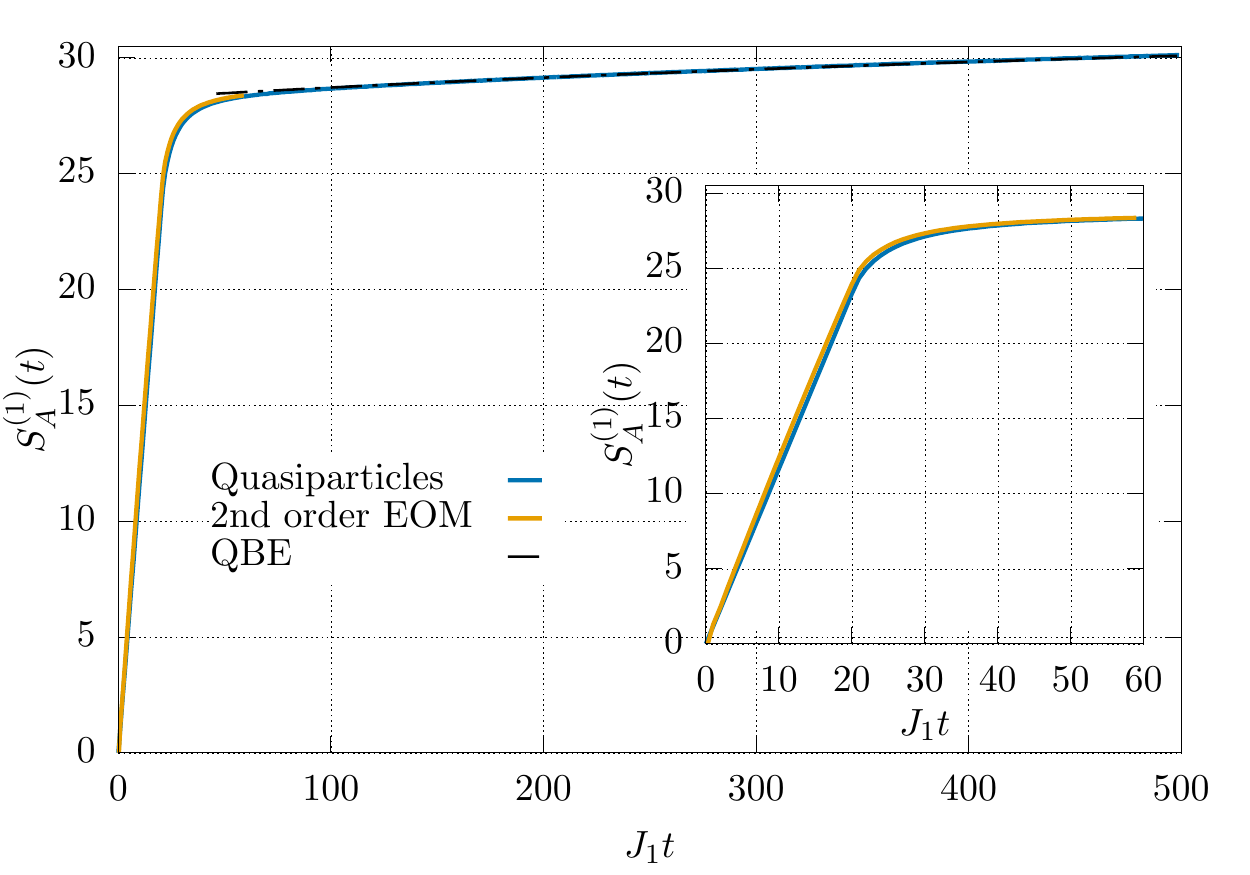} \includegraphics[width=0.45\textwidth]{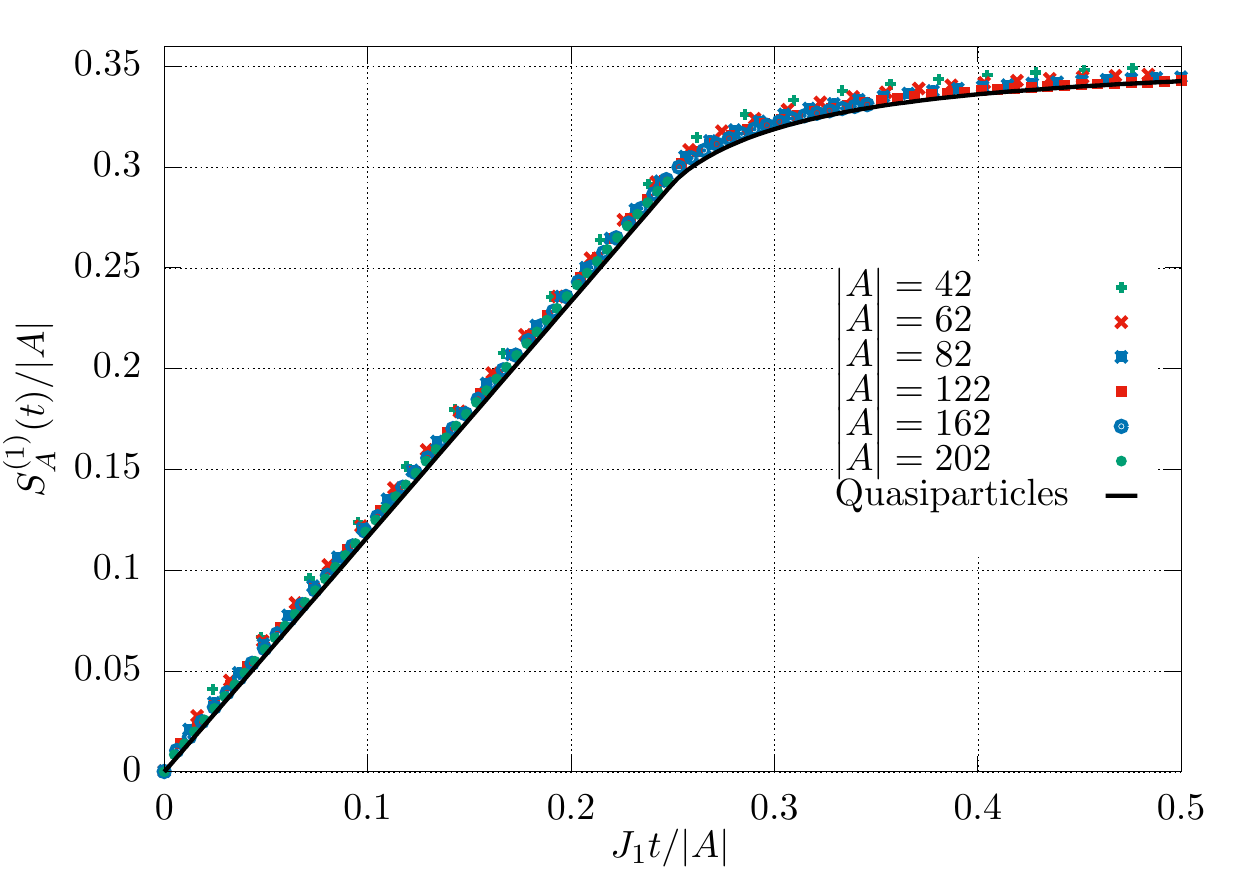} 
\caption{Right. Comparison between the solution of the second order EOM \eqref{Eq:EOM} (orange), the Quantum Boltzmann equation~\eqref{Eq:BB} (dashed-dotted line), and the prediction of the ``scattering quasiparticle" picture~\eqref{eq:quasiparticlesBolt} (blue). The pictures report the behaviour of the entanglement entropy $S_A^{(1)}$ (von Neumann) of a subsystem of size $|A|=120$ after a quench \eqref{eq:quench} with parameters $J_{20}=0$, $\delta_0=0.8$, $U_0=0$, $J_{2}=0.5$, $\delta=0$, $U=0.05$. Left. Collapse of the rescaled von Neumann entropy computed via first order equations for different subsystems $|A|=42,62,82,102$ (points) compared with \eqref{eq:quasiparticlesBolt} (blue). The collapse in the first order data is attained for $\ell_{\rm min} \approx 100$, while $\bar v U^{-2}\approx 400$.} 
\label{Fig:DecayingQP}
\end{figure*}

In the case (i) the effects of the interactions are negligible for all the time needed by $A$ to relax to the deformed GGE. Accordingly, we expect the slopes of the entanglement entropies to be described by the dressed quasiparticle prediction~\eqref{eq:dquasiparticles}. The integrability breaking effects become dominant when the subsystem has already relaxed to the deformed GGE and cause a slow drift towards the thermal state. At the leading order in $U$ such a drift is described by the QBE~\eqref{Eq:BB}. Since during the drift the state is quasi-stationary we expect that for $t\sim U^{-2}$ the entropies can be computed as  
\begin{multline}
S^{(\alpha)}_A(t\sim U^{-2}) = |A| \int_{0}^\pi \!\!\frac{{\rm d} k}{2} (s_{\alpha}[n_{++}(k,U^2 t)]\\ +s_{\alpha}[n_{--}(k,U^2 t)])+O(U)\,,
\label{eq:quasistationaryRenyi}
\end{multline}
where $\{n_{\pm\pm}(q,\tau)\}$ are obtained by solving the Boltzmann Equation \eqref{Eq:BB}. Eq.~\eqref{eq:quasistationaryRenyi} is just the thermodynamic R\'enyi entropy of a free stationary state with occupation numbers $\{n_{\pm\pm}(k,\tau)\}$. Noting that $\{n_{\pm\pm}(q,\tau)\}$ are almost constant for times $t<|A|/v$ we can combine \eqref{eq:quasistationaryRenyi} with \eqref{eq:dquasiparticles}. In this way we obtain the following quasiparticle prediction valid for all times   
\begin{align}
S^{(\alpha)}_{{\rm Bq}, A}(t)&= \int_{0}^\pi \!\!\frac{{\rm d} k}{2}\, {\rm min}(|v^{\rm dr}_+(k,U^2 t)-v^{\rm dr}_-(k,U^2 t)| t, |A|) \notag\\ 
&\qquad\times \left({s_{\alpha}[n^{\rm dr}_{++}(k,U^2 t)]+s_{\alpha}[n^{\rm dr}_{--}(k,\tau)]}\right)\!\!.
\label{eq:quasiparticlesBolt}
\end{align}
Here $v^{\rm dr}_\pm(k,\tau)$ and $n^{\rm dr}_{++}(k,\tau)$ are obtained by replacing $n_{\pm}(k)$ with $n_{\pm}(k,\tau)$ in \eqref{eq:dressedn} (cf.\ \eqref{eq:dE} and \eqref{eq:dGGE}). In this way \eqref{eq:quasiparticlesBolt} is accurate up to $O(U)$ for $t\sim U^{-1}$ (for small enough $U$ the dressing effects become 
negligible and one can safely use $v_\pm(k)$ and $n_{++}(k,\tau)$ in Eq.~\eqref{eq:quasiparticlesBolt}).  Note that in this regime \eqref{eq:quasiparticlesBolt} agrees with \eqref{eq:dquasiparticles} because 
\be
s_{\alpha}[n^{\rm dr}_{++}(k,0)] = \frac{s_{\alpha}[n^{\rm dr}_{++}(k,0)]+s_{\alpha}[n^{\rm dr}_{--}(k,0)]}{2}\,.
\ee
In the case (ii) the effects of the interactions become significant much before the quasi-relaxation of $A$ to the deformed GGE and hence we do not expect the quasiparticle picture (even the dressed one) to correctly describe the slope of the entanglement entropies. Indeed, the latter completely neglects all other mechanisms for spreading and production of entanglement that are active in the non-integrable regime. Nevertheless for large enough times we still expect the system to relax to a quasi-stationary stated described by the Boltzmann equation and \eqref{eq:quasistationaryRenyi} to apply.

The above considerations imply that the verification of  \eqref{eq:quasiparticlesBolt} by means of the equations of motion \eqref{Eq:EOM} requires some care. One needs to consider $|A|>\ell_{\rm min}$ such that the quasiparticle picture can hold, but, at the same time, always keep $|A|\ll  \bar v  U^{-2}$. An intuitive way to estimate $\ell_{\rm min}$ is to look at the collapse of $S^{(\alpha)}_{A}(t)/|A|$ as a function of $t/|A|$. However, it turns out that --- at fixed values of the parameters --- the values of  $|A|$ at which we observe the collapse depend on the specific EOM used: the solution of the second order EOM~\eqref{Eq:EOM} attains its scaling form much before (i.e. form for much smaller $|A|$) that of the first order EOM \eqref{Eq:firstorderEOM1}, see e.g.\ Fig.~\ref{Fig:DifferentScaling}. Since the first order equations are asymptotically described by the quasiparticle picture, we conjecture that the minimal length is set by the collapse of the latter.

\begin{figure}[t]
\includegraphics[width=0.45\textwidth]{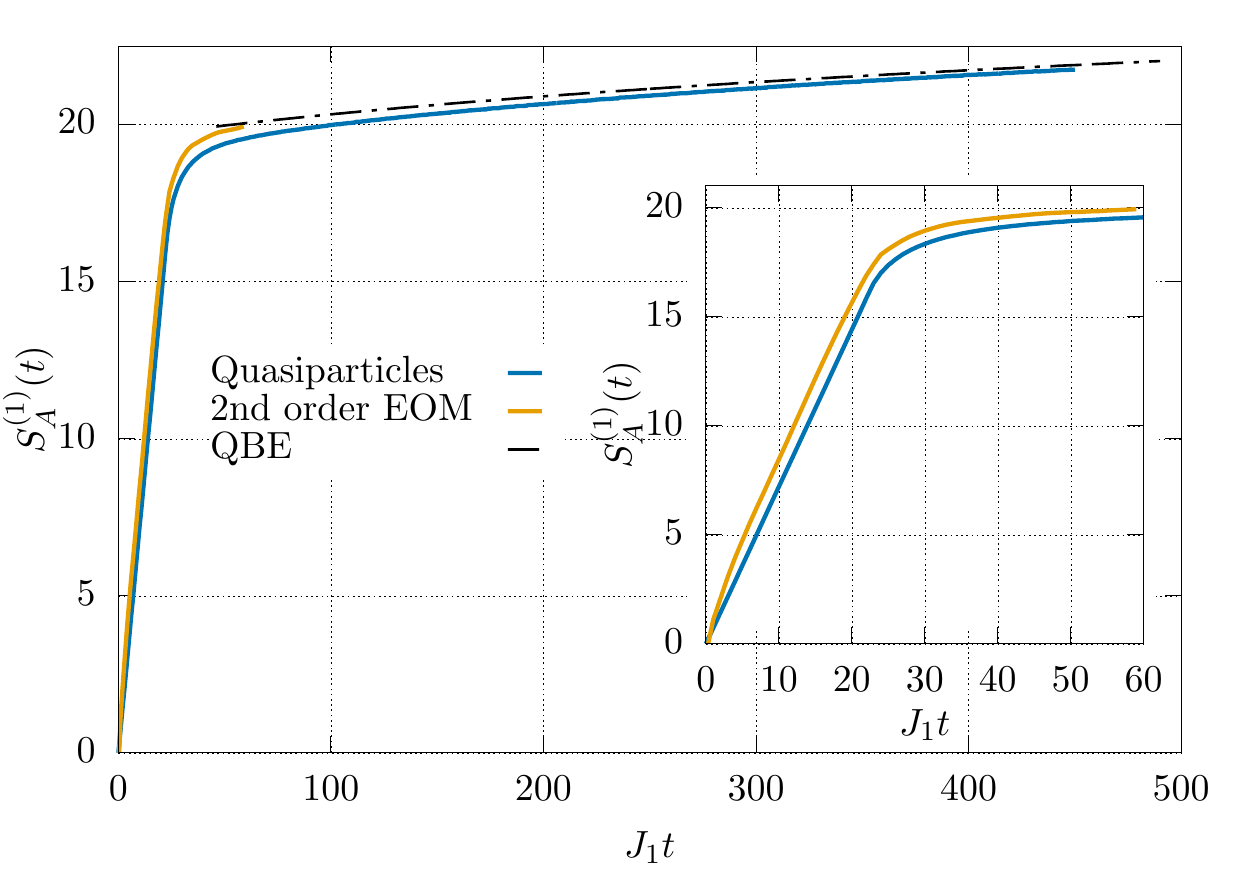}
\caption{Comparison between the solution of the second order EOM \eqref{Eq:EOM} (orange), the Quantum Boltzmann equation~\eqref{Eq:BB} (dashed-dotted line), and the prediction of the ``scattering quasiparticle" picture~\eqref{eq:quasiparticlesBolt} (blue). The pictures report the behaviour of the entanglement entropy $S_A^{(1)}$ (von Neumann) of a subsystem of size $|A|=120$ after a quench \eqref{eq:quench} with parameters $J_{20}=0$, $\delta_0=0.8$, $U_0=0$, $J_{2}=0.5$, $\delta=0.1$, $U=0.1$. The collapse in the first order data is attained for $\ell_{\rm min} \approx 200$, while $\bar v U^{-2}\approx 100$.} 
\label{Fig:DecayingQP2}
\end{figure}

Identifying in this way the regime (ii) we find that \eqref{eq:quasiparticlesBolt} shows a satisfactory agreement with the results of EOM and QBE --- for a representative example see Fig.~\ref{Fig:DecayingQP}. As expected, however, for ${|A|> \bar v U^{-2}}$ Eq.~\eqref{eq:quasiparticlesBolt} describes quantitatively only the late time regime ($t\sim U^{-2}$), see Fig~\ref{Fig:DecayingQP2}. Interestingly the behaviour reported in Fig~\ref{Fig:DecayingQP2} appears to be general: in the initial and intermediate time regime the quasiparticle picture gives a lower bound for the entanglement growth. This can be understood by imagining that together with the entanglement growth due to the spreading quasiparticles there is a further increase of the entanglement due to integrability-breaking effects. 

We conclude this section by recalling that the quasiparticle approach we developed is expected to work exactly in the same manner every time that the unperturbed model is 
described by free particles (and can be also used for more complicated entanglement measures, such as the negativity \cite{ac-18b}). 
Instead, when the unperturbed model is an interacting integrable one, it is only known how to adapt the quasiparticle picture to the time evolution of the von 
Neumann entropy \cite{CA,CA2}, while for R\'enyi entropies there are still open issues, see e.g. Refs. \cite{ac-17a,ac-17b,mac-18,ABF19}.

\section{Conclusions}
\label{sec:conclusions}

In this paper we studied the spreading and generation of entanglement in a weakly interacting system of lattice fermions using Equations of Motion techniques~\cite{SK:EOM,INJPhys,NessiArxiv15,BEGR:PRL,ESY:QBE,LS:QBE, BEGR:PRL, BEGR:PRB, Bonitzbook, FMS:NIQBE, FMS:QBE, KBbook}. We found that for small enough interactions --- parametrised by their strength $U$ --- the entanglement entropies show the typical prethermalization behaviour~\cite{MK:prethermalization}: they first approach a quasi-stationary plateau described by a deformed GGE and then, on a separate timescale $\tau_{\rm th}\sim U^{-2}$, they start relaxing towards their thermal value. This behaviour has been interpreted by means of a modified quasiparticle picture where the contribution of each pair to the entanglement --- normally time-independent --- depends on $U^2 t$ and is obtained by solving a Quantum Boltzmann Equation. This modified quasiparticle picture predicts the correct quantitative behaviour of the entanglement entropies of the subsystem $A$ whenever $\tau_A$ --- the timescale over which $A$ relaxes to the deformed GGE --- is much smaller than $\tau_{\rm th}$ --- the timescale associated with integrability breaking. In the opposite case it describes quantitatively only the late-time regime while it underestimates the slope of the entanglement entropies.

There are two immediate future directions for the research presented in our work. First it would be interesting to generalise our findings to the case of weak perturbations to strongly interacting integrable systems, combining our modified quasiparticle picture with the recent results~\cite{LWGV:20, DBD:20, FGV:19} on the Quantum Boltzmann Equation for interacting integrable models. Second it would be interesting to search for a simple description of the integrabilty-breaking correction that we observed in the slope of the entanglement entropies for $\tau_A> \tau_{\rm th}$. Note that similar integrability-breaking effects are also visible in equal-time two-point functions for large enough separation of the two points.

\section*{Acknowledgments}
BB acknowledges support by the EU Horizon 2020 program through the ERC Advanced Grant OMNES No. 694544, and by the Slovenian Research Agency (ARRS) under the Programme P1-0402. 
PC acknowledges support from ERC under Consolidator grant number 771536 (NEMO).



\begin{thebibliography}{999}


\bibitem{PolkonikovRMP11}
A.~Polkovnikov, K.~Sengupta, A.~Silva, and M.~Vengalattore,
\textit{Colloquium: Nonequilibrium dynamics of closed interacting quantum systems}, 
\href{http://dx.doi.org/10.1103/RevModPhys.83.863}{Rev. Mod. Phys. {\bf 83}, 863 (2011)}.

\bibitem{GE15}
C.~Gogolin and J.~Eisert, 
\textit{Equilibration, thermalisation, and the emergence of statistical mechanics in closed quantum systems}, 
\href{http://iopscience.iop.org/article/10.1088/0034-4885/79/5/056001}{Rep. Prog. Phys. {\bf 79}, 056001 (2016)}. 

\bibitem{DKPR15}
L.~D'Alessio, Y.~Kafri, A.~Polkovnikov, and M.~Rigol,
\textit{From Quantum Chaos and Eigenstate Thermalization to Statistical Mechanics and Thermodynamics}, 
\href{http://dx.doi.org/10.1080/00018732.2016.1198134}{Adv. Phys. {\bf 65}, 239 (2016)}.

\bibitem{SI}
P. Calabrese, F. Essler, and G. Mussardo, \href{http://iopscience.iop.org/article/10.1088/1742-5468/2016/06/064001}{J. Stat. Mech. (2016) 064001}. 












\bibitem{C:18}
P. Calabrese, \emph{Entanglement and thermodynamics in non-equilibrium isolated quantum systems}, 
\href{https://doi.org/10.1016/j.physa.2017.10.011}{Physica A {\bf 504}, 31 (2018)}.

\bibitem{DLS:13}
J. M. Deutsch, H. Li, and A. Sharma, \emph{Microscopic origin of thermodynamic entropy in isolated systems}, \href{https://doi.org/10.1103/PhysRevE.87.042135}{Phys. Rev. E {\bf 87}, 042135 (2013)}.

\bibitem{BAH:15}
W. Beugeling, A. Andreanov, and M. Haque, \emph{Global characteristics of all eigenstates of local many-body Hamiltonians: participation ratio and entanglement entropy}, \href{https://doi.org/10.1088/1742-5468/2015/02/P02002}{J. Stat. Mech. (2015) P02002}.

\bibitem{G:14}
V. Gurarie, \emph{Global large time dynamics and the generalized Gibbs ensemble}, \href{
https://doi.org/10.1088/1742-5468/2013/02/P02014}{J. Stat. Mech. (2013) P02014}.

\bibitem{SPR:11}
L. F. Santos, A. Polkovnikov, and M. Rigol, \emph{Entropy of Isolated Quantum Systems after a Quench}, \href{https://doi.org/10.1103/PhysRevLett.107.040601}{Phys. Rev. Lett. {\bf 107}, 040601 (2011)}.

\bibitem{ckc-14}
M.~Collura, M.~Kormos, and P.~Calabrese, {\it Stationary entropies following an interaction quench in $1D$ Bose gas}, 
\href{https://doi.org/10.1088/1742-5468/2014/01/P01009}{J. Stat. Mech. P01009 (2014)}.

\bibitem{kaufman-2016}
A.~M.~Kaufman, M.~E.~Tai, A.~Lukin, M.~Rispoli, R.~Schittko, P.~M.~Preiss, and  M.~Greiner, 
Quantum thermalisation through entanglement in an isolated many-body system, 
\href{http://dx.doi.org/10.1126/science.aaf6725}{Science {\bf 353}, 794  (2016)}.

\bibitem{ABGH:15}
C. T. Asplund, A. Bernamonti, F. Galli, and T. Hartman, \emph{Entanglement Scrambling in 2D Conformal Field Theory}, \href{https://doi.org/10.1007/JHEP09(2015)110}{J. High Energy Phys. {\bf 09} (2015) 110}.

\bibitem{HRT:15}
V. Hubeny, M. Rangamani, and T. Takayanagi, \emph{A Covariant Holographic Entanglement Entropy Proposal}, \href{https://doi.org/10.1088/1126-6708/2007/07/062}{J. High Energy Phys. 07 (2007) 062.}

\bibitem{AAAL:10} 
J. Abajo-Arrastia, J. Apar\'icio, and E. L\'opez, \emph{Holographic Evolution of Entanglement Entropy}, \href{
https://doi.org/10.1007/JHEP11(2010)149}{J. High Energy Phys. 11 (2010) 149.}

\bibitem{HM:13}
T. Hartman and J. Maldacena, \emph{Time Evolution of Entanglement Entropy from Black Hole Interiors}, \href{
https://doi.org/10.1007/JHEP05(2013)014}{J. High Energy Phys. 05 (2013) 014.}

\bibitem{LS:14}
H. Liu and S. J. Suh, \emph{Entanglement Tsunami: Universal Scaling in Holographic Thermalization}, \href{https://doi.org/10.1103/PhysRevLett.112.011601}{Phys. Rev. Lett. 112, 011601 (2014).}

\bibitem{LM:CFT}
S. Leichenauer and M. Moosa, \emph{Entanglement tsunami in (1+1)-dimensions}, \href{https://doi.org/10.1103/PhysRevD.92.126004}{Phys. Rev. D {\bf 92}, 126004 (2015)}.


\bibitem{CLM:16}
H. Casini, H. Liu, and M. Mezei, \emph{Spread of Entanglement and Causality}, \href{
https://doi.org/10.1007/JHEP07(2016)077}{J. High Energy Phys. {\bf 07} (2016) 077}.





\bibitem{HP:07}
P. Hayden and J. Preskill, \emph{Black holes as mirrors: quantum information in random subsystems}, \href{https://doi.org/10.1088/1126-6708/2007/09/120}{J. High Energy Phys. {\bf 2007}, 120 (2007)}.


\bibitem{SS:08}
Y. Sekino and L. Susskind, \emph{Fast scramblers}, \href{https://doi.org/10.1088/1126-6708/2008/10/065}{J. High Energy Phys. {\bf 2008}, 065 (2008)}.


\bibitem{AC:19}
V. Alba and P. Calabrese, \emph{Quantum information scrambling after a quantum quench}, \href{https://doi.org/10.1103/PhysRevB.100.115150}{Phys. Rev. B {\bf 100}, 115150 (2019)}.

\bibitem{mac-20}
R. Modak, V. Alba, and P. Calabrese, {\it Entanglement revivals as a probe of scrambling in finite quantum systems}, 
\href{https://arxiv.org/abs/2004.08706}{arxiv:2004.08706}.


\bibitem{SWVC:PRL}
N. Schuch, M. M. Wolf, F. Verstraete, and J. I. Cirac, {\it Entropy Scaling and Simulability by Matrix Product States},
\href{https://doi.org/10.1103/PhysRevLett.100.030504}{Phys. Rev. Lett. {\bf 100}, 030504 (2008)}.

\bibitem{SWVC:NJP}
N. Schuch, M. M. Wolf, K. G. H. Vollbrecht, and J. I. Cirac, {\it On entropy growth and the hardness of simulating time evolution},
\href{https://doi.org/10.1088/1367-2630/10/3/033032}{New J. Phys. {\bf 10}, 033032 (2008)}.

\bibitem{PV:08}
A. Perales and G. Vidal, {\it Entanglement growth and simulation efficiency in one-dimensional quantum lattice systems},
\href{https://doi.org/10.1103/PhysRevA.78.042337}{Phys. Rev. A {\bf 78}, 042337 (2008)}.

\bibitem{HCTDL:12}
P. Hauke, F. M. Cucchietti, L. Tagliacozzo, I. Deutsch, and M. Lewenstein, {\it Can one trust quantum simulators?}
\href{https://doi.org/10.1088/0034-4885/75/8/082401}{Prog. Phys. {\bf 75} 082401 (2012)}.

\bibitem{D:17}
J. Dubail, \emph{Entanglement scaling of operators: a conformal field theory approach, with a glimpse of simulability of long-time dynamics in 1+1d}, \href{https://doi.org/10.1088/1751-8121/aa6f38}{J. Phys. A {\bf 50}, 234001 (2017)}.


\bibitem{CC}
P. Calabrese and J. Cardy, \emph{Evolution of Entanglement Entropy in One-Dimensional Systems}, \href{http://dx.doi.org/10.1088/1742-5468/2005/04/P04010}{J. Stat. Mech. (2005) P04010}.


\bibitem{FC:exactXY} M.~Fagotti and P.~Calabrese, \emph{Evolution of entanglement entropy following a quantum quench: Analytic results for the XY chain in a transverse magnetic field}, \href{https://doi.org/10.1103/PhysRevA.78.010306}{Phys. Rev. A {\bf 78}, 010306(R)}.

\bibitem{ep-08}
V. Eisler and I. Peschel, \emph{Entanglement in a periodic quench}, \href{https://doi.org/10.1002/andp.200810299}{Ann. Phys. (Berlin) {\bf 17}, 410 (2008)}.

\bibitem{nr-14}
M. G.~Nezhadhaghighi and M. A. Rajabpour, \emph{Entanglement dynamics in short- and long-range harmonic oscillators}, \href{https://doi.org/10.1103/PhysRevB.90.205438}{Phys. Rev. B {\bf 90}, 205438 (2014)}.

\bibitem{bkc-14}
L. Bucciantini, M. Kormos, and P. Calabrese, \emph{Quantum quenches from excited states in the Ising chain}, \href{https://doi.org/10.1088/1751-8113/47/17/175002}{J. Phys. A {\bf 47}, 175002 (2014)}.


\bibitem{bhy-17}
E. Bianchi, L. Hackl, and N. Yokomizo, \emph{Linear growth of the entanglement entropy and the Kolmogorov-Sinai rate}, 
\href{https://doi.org/10.1007/JHEP03(2018)025}{J. High Energ. Phys. (2018) 2018: 25}.

\bibitem{hbmr-17}
L. Hackl, E. Bianchi, R. Modak, and M. Rigol, \emph{Entanglement production in bosonic systems: Linear and logarithmic growth}, \href{https://doi.org/10.1103/PhysRevA.97.032321}{Phys. Rev. A {\bf 97}, 032321 (2018)}.


\bibitem{buyskikh-2016}
A. S. Buyskikh, M. Fagotti, J. Schachenmayer, F. Essler, and A. J. Daley, \emph{Entanglement growth and correlation spreading with variable-range interactions in spin and fermionic tunneling models}, \href{http://dx.doi.org/10.1103/PhysRevA.93.053620}{Phys.\ Rev.\ A {\bf 93}, 053620 (2016)}. 


\bibitem{cotler-2016}
J.~S.~Cotler, M.~P.~Hertzberg, M.~Mezei, and M.~T.~Mueller, \emph{Entanglement growth after a global quench in free scalar field theory}, \href{https://doi.org/10.1007/JHEP11(2016)166}{JHEP {\bf 11}, 166 (2016)}. 

\bibitem{triplets}
B. Bertini, E. Tartaglia, and P. Calabrese, \emph{Entanglement and diagonal entropies after a quench with no pair structure}, \href{http://dx.doi.org/10.1088/1742-5468/aac73f}{J. Stat. Mech. (2018) 063104}. 


\bibitem{intertwined}
A. Bastianello and P. Calabrese, \emph{Spreading of entanglement and correlations after a quench with intertwined quasiparticles}, \href{http://dx.doi.org/10.21468/SciPostPhys.5.4.033}{SciPost Phys. {\bf 5}, 033 (2018)}.

\bibitem{BFPC18}
B.~Bertini, M.~Fagotti, L.~Piroli, and P.~Calabrese, \emph{Entanglement evolution and generalised hydrodynamics: noninteracting systems}, \href{http://dx.doi.org/10.1088/1751-8121/aad82e}{J. Phys. A: Math. Theor. {\bf 51}, 39LT01 (2018)}.



\bibitem{CA} 
V.~Alba and P.~Calabrese, \emph{Entanglement and thermodynamics after a quantum quench in integrable systems}, 
\href{https://doi.org/10.1073/pnas.1703516114}{PNAS {\bf 114}, 7947 (2017)}.


\bibitem{CA2}
V. Alba and P. Calabrese, \emph{Entanglement dynamics after quantum quenches in generic integrable systems}, \href{https://doi.org/10.21468/SciPostPhys.4.3.017}{SciPost Phys. {\bf 4}, 017 (2018)}.

\bibitem{mbpc-17}
M. Mestyan, B. Bertini, L. Piroli, and P. Calabrese, \emph{Exact solution for the quench dynamics of a nested integrable system}, \href{http://dx.doi.org/10.1088/1742-5468/aa7df0}{J. Stat. Mech. (2017) 083103}.

\bibitem{mkz-17}
C. P. Moca, M. Kormos, and G. Zarand, \emph{Hybrid Semiclassical Theory of Quantum Quenches in One-Dimensional Systems}, \href{http://dx.doi.org/10.1103/PhysRevLett.119.100603}{Phys. Rev. Lett. {\bf 119}, 100603 (2017)}.


\bibitem{alba-inh}
V.~Alba, \emph{Entanglement and quantum transport in integrable systems}, \href{http://dx.doi.org/10.1103/PhysRevB.97.245135}{Phys. Rev. B {\bf 97}, 245135 (2018)}. 

\bibitem{ABF19}
V. Alba, B. Bertini, and M. Fagotti, \emph{Entanglement evolution and generalised hydrodynamics: interacting integrable systems}, \href{http://dx.doi.org/10.21468/SciPostPhys.7.1.005}{SciPost Phys. {\bf 7}, 005 (2019)}. 



\bibitem{lauchli-2008}
A. Laeuchli and C. Kollath, \emph{Spreading of correlations and entanglement after a quench in the one-dimensional Bose-Hubbard model}, \href{https://doi.org/10.1088/1742-5468/2008/05/P05018}{J. Stat. Mech.  P05018 (2008)}.

\bibitem{KH:NonIntEnt} 
H. Kim and D. A. Huse, \emph{Ballistic Spreading of Entanglement in a Diffusive Nonintegrable System}, \href{https://doi.org/10.1103/PhysRevLett.111.127205}{Phys. Rev. Lett. {\bf 111}, 127205 (2013)}.
	
\bibitem{ckt-18}
M.~Collura, M.~Kormos, and G.~Takacs, \emph{Dynamical manifestation of the Gibbs paradox after a quantum quench}, \href{http://dx.doi.org/10.1103/PhysRevA.98.053610}{Phys. Rev. A {\bf 98}, 053610 (2018)}.

\bibitem{FNR:longrangehigherd}
I. Frerot, P. Naldesi, and T. Roscilde, \emph{Multispeed Prethermalization in Quantum Spin Models with Power-Law Decaying Interactions}, \href{https://doi.org/10.1103/PhysRevLett.120.050401}{Phys. Rev. Lett. {\bf 120}, 050401 (2018)}.

\bibitem{PL:kickedIsing} 
R. Pal and A. Lakshminarayan, \emph{Entangling power of time-evolution operators in integrable and nonintegrable many-body systems}, \href{https://doi.org/10.1103/PhysRevB.98.174304}{Phys. Rev. B {\bf 98}, 174304 (2018)}.

\bibitem{BKP:entropy}
B. Bertini, P. Kos, and T. Prosen, \emph{Entanglement Spreading in a Minimal Model of Maximal Many-Body Quantum Chaos},  \href{https://doi.org/10.1103/PhysRevX.9.021033}{Phys. Rev. X {\bf 9}, 021033 (2019)}.

\bibitem{GoLa19} 
S. Gopalakrishnan and A. Lamacraft, \emph{Unitary circuits of finite depth and infinite width from quantum channels},
\href{http://dx.doi.org/10.1103/PhysRevB.100.064309}{Phys. Rev. B {\bf 100}, 064309 (2019)}.

\bibitem{brydges-2018}
T.~Brydges, A.~Elben, P.~Jurcevic, B.~Vermersch, C.~Maier, B.~P. Lanyon,  P. Zoller, R. Blatt, and C. F. Roos,
  \textit{Probing entanglement entropy via randomized measurements},
 \href{http://dx.doi.org/10.1126/science.aau4963}{Science {\bf 364}, 260 (2019)}, 

\bibitem{PBCP20} 
L. Piroli, B. Bertini, J. I. Cirac, and T. Prosen, \emph{Exact dynamics in dual-unitary quantum circuits},
\href{http://dx.doi.org/10.1103/PhysRevB.101.094304}{Phys. Rev. B {\bf 101}, 094304 (2020)}.

\bibitem{smg-20}
F. M. Surace, P. P. Mazza, G. Giudici, A. Lerose, A. Gambassi, and M. Dalmonte, {\it Lattice gauge theories and string dynamics in Rydberg atom quantum simulators},
\href{http://dx.doi.org/10.1103/PhysRevX.10.021041}{Phys. Rev. X 10, 021041 (2020)}.


\bibitem{DMCF:06}
G. De Chiara, S. Montangero, P. Calabrese and R. Fazio, \emph{Entanglement entropy dynamics of Heisenberg chains}, \href{https://doi.org/10.1088/1742-5468/2006/03/P03001}{J. Stat. Mech. (2006) P03001}.

\bibitem{ZPP:08}
M. Znidari\v c, T. Prosen, and P. Prelov\v sek, \emph{Many-body localization in the Heisenberg XXZ magnet in a random field}, \href{https://doi.org/10.1103/PhysRevB.77.064426}{Phys. Rev. B {\bf 77}, 064426 (2008)}.

\bibitem{bpm-12}
J. H. Bardarson, F. Pollmann, and J. E. Moore,
{\it Unbounded growth of entanglement in models of many-body localization},
\href{http://dx.doi.org/10.1103/PhysRevLett.109.017202}{Phys. Rev. Lett. {\bf 109}, 017202 (2012)}.

\bibitem{ModulatedIsing}
F.~Igl\'oi, G.~Ro\'osz, and Y.-C.~Lin, \emph{Non-equilibrium quench dynamics in quantum quasicrystals}, \href{https://doi.org/10.1088/1367-2630/15/2/023036}{New J. Phys. {\bf 15} 023036 (2013)}.

\bibitem{ModulatedIsing2}
G.~Ro\'osz, U.~Divakaran, H.~Rieger, and F.~Igl\'oi, \emph{Nonequilibrium quantum relaxation across a localization-delocalization transition}, \href{https://doi.org/10.1103/PhysRevB.90.184202}{Phys. Rev. B {\bf 90}, 184202 (2014)}.


\bibitem{Vosk2014}
R. Vosk and E. Altman, 
\emph{Dynamical Quantum Phase Transitions in Random Spin Chains},
\href{http://dx.doi.org/10.1103/PhysRevLett.112.217204}{Phys. Rev. Lett. {\bf 112}, 217204 (2014)}. 

\bibitem{NH:MBL}
R. Nandkishore and D. A. Huse, \emph{Many-Body Localization and Thermalization in Quantum Statistical Mechanics}, \href{https://doi.org/10.1146/annurev-conmatphys-031214-014726}{Ann. Rev. Condens. Matter Phys. {\bf 6}, 15 (2015)}.

\bibitem{isl-12}
F. Igloi, Z. Szatmari, and Y.-C. Lin, {\it Entanglement entropy dynamics of disordered quantum spin chains}, 
\href{https://doi.org/10.1103/PhysRevB.85.094417}{Phys. Rev. B {\bf 85}, 094417 (2012)}

\bibitem{NRH:18}
A. Nahum, J. Ruhman, and D. A. Huse, \emph{Dynamics of entanglement and transport in one-dimensional systems with quenched randomness} \href{https://doi.org/10.1103/PhysRevB.98.035118}{Phys. Rev. B {\bf 98}, 035118 (2018)}.

\bibitem{KCTC:17}
M. Kormos, M. Collura, G. Tak\'acs, and P. Calabrese, \emph{Real-time confinement following a quantum quench to a non-integrable model}, \href{https://doi.org/10.1038/nphys3934}{Nature Physics {\bf 13}, 246 (2017)}.

\bibitem{jkr-19}
A. J. A. James, R. M. Konik, and N. J. Robinson,
{\it Nonthermal States Arising from Confinement in One and Two Dimensions},
\href{https://doi.org/10.1103/PhysRevLett.122.130603}{Phys. Rev. Lett. {\bf 122}, 130603 (2019)}.

\bibitem{clsv-20}
O. A. Castro-Alvaredo, M. Lencses, I. M. Szecsenyi, and J. Viti, {\it Entanglement Oscillations near a Quantum Critical Point}, 
\href{https://doi.org/10.1103/PhysRevLett.124.230601}{Phys. Rev. Lett. {\bf 124}, 230601 (2020)}.


\bibitem{lsmpcg}
A. Lerose, F. M. Surace, P. P. Mazza, G. Perfetto, M. Collura, and A. Gambassi,
{\it Quasilocalized dynamics from confinement of quantum excitations},
\href{https://arxiv.org/abs/1911.07877}{arXiv:1911.07877}.

\bibitem{Schwinger}
T. Chanda, J. Zakrzewski, M. Lewenstein, and L. Tagliacozzo, \emph{Confinement and Lack of Thermalization after Quenches in the Bosonic Schwinger Model}, \href{https://doi.org/10.1103/PhysRevLett.124.180602}{Phys. Rev. Lett. {\bf 124}, 180602 (2020)}. 

\bibitem{scar1}
C. J. Turner, A. A. Michailidis, D. A. Abanin, M. Serbyn, and Z. Papic, {\it Quantum many-body scars},
\href{https://doi.org/10.1038/s41567-018-0137-5}{Nature Phys. {\bf 14},  745 (2018)}.

\bibitem{scar2}
S. Choi, C. J. Turner, H. Pichler, W. W. Ho, A. A. Michailidis, Z. Papic, M. Serbyn, M. D. Lukin, and D. A. Abanin,
{\it Emergent SU(2) dynamics and perfect quantum many-body scars},
\href{https://doi.org/10.1103/PhysRevLett.122.220603}{Phys. Rev. Lett. {\bf 122}, 220603 (2019)}.

\bibitem{NRVH:17}
A. Nahum, J. Ruhman, S. Vijay, and J. Haah, \emph{Quantum Entanglement Growth under Random Unitary Dynamics}, \href{https://doi.org/10.1103/PhysRevX.7.031016}{Phys. Rev. X {\bf 7}, 031016 (2017)}. 

\bibitem{Nahum:operatorspreadingRU}
A. Nahum, S. Vijay, and J. Haah, \emph{Operator Spreading in Random Unitary Circuits}, \href{https://doi.org/10.1103/PhysRevX.8.021014}{Phys. Rev. X {\bf 8}, 021014 (2018)}.

\bibitem{ZN:nonrandommembrane}
T. Zhou and A. Nahum, \emph{The entanglement membrane in chaotic many-body systems}, 
\href{https://arxiv.org/abs/1912.12311}{arXiv:1912.12311 (2019)}. 





\bibitem{MK:prethermalization}
M. Moeckel and S. Kehrein, 
\textit{Interaction Quench in the Hubbard Model},
\href{http://dx.doi.org/10.1103/PhysRevLett.100.175702}{Phys. Rev. Lett. {\bf 100}, 175702 (2008)};  
\textit{Real-time evolution for weak interaction quenches in quantum systems},
\href{http://dx.doi.org/10.1016/j.aop.2009.03.009}{Ann. Phys. {\bf 324}, 2146 (2009)}.

\bibitem{RoschPRL08}
A.~Rosch, D.~Rasch, B.~Binz, and M.~Vojta, 
\textit{Metastable Superfluidity of Repulsive Fermionic Atoms in Optical Lattices}, 
\href{http://dx.doi.org/10.1103/PhysRevLett.101.265301}{Phys. Rev. Lett. {\bf 101}, 265301 (2008)}.

\bibitem{KollarPRB11}
M.~Kollar, F. A.~Wolf, and M.~Eckstein,
\textit{Generalized Gibbs ensemble prediction of prethermalization plateaus and their relation to nonthermal steady states in integrable systems},   
\href{http://dx.doi.org/10.1103/PhysRevB.84.054304}{Phys. Rev. B {\bf 84}, 054304 (2011)}.

\bibitem{worm13}
M.~van~den~Worm, B.~C.~Sawyer, J.~J.~Bollinger, and M.~Kastner, 
\textit{Relaxation timescales and decay of correlations in a long-range interacting quantum simulator}, 
\href{http://dx.doi.org/10.1088/1367-2630/15/8/083007}{New J. Phys. {\bf 15}, 083007 (2013)}.

\bibitem{MarcuzziPRL13}
M.~Marcuzzi, J.~Marino, A.~Gambassi, and A.~Silva, 
\textit{Prethermalization in a Nonintegrable Quantum Spin Chain after a Quench}, 
\href{http://dx.doi.org/10.1103/PhysRevLett.111.197203}{Phys. Rev. Lett. {\bf 111}, 197203 (2013)}.

\bibitem{EsslerPRB14}
F.~H.~L.~Essler, S.~Kehrein, S.~R.~Manmana, and N.~J.~Robinson, 
\textit{Quench dynamics in a model with tuneable integrability breaking}, 
\href{http://dx.doi.org/10.1103/PhysRevB.89.165104}{Phys. Rev. B {\bf 89}, 165104 (2014)}.

\bibitem{NIC14}
N.~Nessi, A.~Iucci and M.~A.~Cazalilla, 
\textit{Quantum Quench and Prethermalization Dynamics in a Two-Dimensional Fermi Gas with Long-Range Interactions}, 
\href{http://dx.doi.org/10.1103/PhysRevLett.113.210402}{Phys. Rev. Lett. {\bf 113}, 210402 (2014)}.

\bibitem{Fagotti14}
M. Fagotti,
\textit{On conservation laws, relaxation and pre-relaxation after a quantum quench}, 
\href{http://dx.doi.org/10.1088/1742-5468/2014/03/P03016}{J. Stat. Mech. (2014) P03016}.

\bibitem{konik14}
G.~P.~Brandino, J.-S.~Caux, and R.~M.~Konik, 
\textit{Glimmers of a Quantum KAM Theorem: Insights from Quantum Quenches in One-Dimensional Bose Gases}, 
\href{http://dx.doi.org/10.1103/PhysRevX.5.041043}{Phys. Rev. X {\bf 5}, 041043 (2015)}. 

\bibitem{BF15}
B. Bertini and M. Fagotti, 
\textit{Pre-relaxation in weakly interacting models},
\href{http://dx.doi.org/10.1088/1742-5468/2015/07/P07012}{J. Stat. Mech. (2015) P07012}.

\bibitem{CTGM:pret} 
A.~Chiocchetta, M.~Tavora, A.~Gambassi, and A.~Mitra, 
\emph{Short-time universal scaling in an isolated quantum system after a quench},
\href{http://dx.doi.org/10.1103/PhysRevB.91.220302}{Phys. Rev. B {\bf 91}, 220302(R) (2015)}; \href{http://dx.doi.org/10.1103/PhysRevB.92.219901}{Phys. Rev. {\bf B} 92, 219901(E) (2015).}

\bibitem{knap15}
M. Babadi, E.~Demler, and M.~Knap, 
\textit{Far-from-Equilibrium Field Theory of Many-Body Quantum Spin Systems: Prethermalization and Relaxation of Spin Spiral States in Three Dimensions}, 
\href{http://dx.doi.org/10.1103/PhysRevX.5.041005}{Phys. Rev. X {\bf 5}, 041005 (2015)}.

\bibitem{SmacchiaPRB15}
P.~Smacchia, M.~Knap, E.~Demler, and A.~Silva, 
\textit{Exploring dynamical phase transitions and prethermalization with quantum noise of excitations}, 
\href{http://link.aps.org/doi/10.1103/PhysRevB.91.205136}{Phys. Rev. B {\bf 91}, 205136 (2015).}

\bibitem{BEGR:PRL}
B. Bertini, F.~H.~L.~Essler, S.~Groha, and N.~J.~Robinson, 
\textit{Prethermalization and Thermalization in Models with Weak Integrability Breaking}, 
\href{http://doi.org/10.1103/PhysRevLett.115.180601}{Phys. Rev. Lett. {\bf 115}, 180601 (2015)}.

\bibitem{BEGR:PRB}
B. Bertini, F.~H.~L.~Essler, S.~Groha, and N.~J.~Robinson, 
\textit{Thermalization and light cones in a model with weak integrability breaking}, 
\href{http://dx.doi.org/10.1103/PhysRevB.94.245117}{Phys. Rev. B {\bf 94}, 245117 (2016)}. 

\bibitem{FC15}
M. Fagotti and M.~Collura, 
\textit{Universal prethermalization dynamics of entanglement entropies after a global quench}, 
\href{http://arxiv.org/abs/1507.02678}{arXiv:1507.02678 (2015)}.

\bibitem{MenegozJStatMech15}
G.~Menegoz and A.~Silva,
\textit{Prethermalization of weakly interacting bosons after a sudden interaction quench}, 
\href{http://stacks.iop.org/1742-5468/2015/i=5/a=P05035}{J. Stat. Mech. (2015) P05035}.

\bibitem{KaminishiNatPhys15}
E.~Kaminishi, T.~Mori, T.~Ikeda, N.~Tatsuhiko, and M.~Ueda,
\textit{Entanglement pre-thermalization in a one-dimensional Bose gas}, 
\href{http://dx.doi.org/10.1038/nphys3478}{Nat. Phys. {\bf 11}, 1050 (2015)}.

\bibitem{D-14}
G. Delfino, \emph{Quantum quenches with integrable pre-quench dynamics}, \href{http://dx.doi.org/10.1088/1751-8113/47/40/402001}{J. Phys. A  {\bf 47} (2014) 402001}.

\bibitem{dv-17}
G. Delfino and J. Viti, {\it On the theory of quantum quenches in near-critical systems},
\href{http://dx.doi.org/10.1088/1751-8121/aa5660}{J. Phys. A {\bf 50} (2017) 084004}.

\bibitem{AFpret}
V. Alba and M. Fagotti, \href{http://dx.doi.org/10.1103/PhysRevLett.119.010601}{Phys. Rev. Lett. {\bf 119}, 010601 (2017)}

\bibitem{DeRoeck}
K. Mallayya, M. Rigol, W. De Roeck, \emph{Prethermalization and Thermalization in Isolated Quantum Systems}, \href{http://dx.doi.org/10.1103/PhysRevX.9.021027}{Phys. Rev. X {\bf 9}, 021027 (2019)}. 

\bibitem{FGV:19}
A. J. Friedman, S. Gopalakrishnan, R. Vasseur, \emph{Diffusive hydrodynamics from integrability breaking}, \href{http://dx.doi.org/10.1103/PhysRevB.101.180302}{Phys. Rev. B {\bf 101}, 180302 (2020)}. 

\bibitem{DBD:20}
J.~Durnin, M.~J.~Bhaseen, and B.~Doyon, \emph{Non-Equilibrium Dynamics and Weakly Broken Integrability}, \href{https://arxiv.org/abs/2004.11030}{arXiv:2004.11030 (2020)}. 

\bibitem{LWGV:20}
J. Lopez-Piqueres, B. Ware, S. Gopalakrishnan, R. Vasseur, \emph{Hydrodynamics of non-integrable systems from relaxation-time approximation}, \href{https://arxiv.org/abs/2005.13546}{arXiv:2005.13546 (2020)}. 

\bibitem{RFGpret}
P. Ruggiero, L. Foini, and T. Giamarchi, \href{https://arxiv.org/abs/2006.16088}{arXiv:2006.16088 (2020)}.

\bibitem{gring-2012}
M.~Gring, M.~Kuhnert, T.~Langen, T.~Kitagawa, B.~Rauer, M.~Schreitl, I.~Mazets, D.~A.~Smith, E.~Demler, and J.~Schmiedmayer, 
Relaxation Dynamics and Pre-thermalisation in an Isolated Quantum System,
\href{http://dx.doi.org/10.1126/science.1224953}{Science {\bf 337}, 1318 (2012)}.

\bibitem{LangenReview}
T.~Langen, T.~Gasenzer, and J. Schmiedmayer,
\textit{Prethermalization and universal dynamics in near-integrable quantum systems}, 
\href{http://iopscience.iop.org/article/10.1088/1742-5468/2016/06/064009}{J. Stat. Mech. (2016) 064009}.

\bibitem{Dysprosiumcradle}
Y. Tang, W. Kao, K.-Y. Li, S. Seo, K. Mallayya, M. Rigol, S. Gopalakrishnan, and B. L. Lev, \emph{Thermalization near Integrability in a Dipolar Quantum Newton?s Cradle}, \href{https://doi.org/10.1103/PhysRevX.8.021030}{Phys. Rev. X {\bf 8}, 021030 (2018)}

\bibitem{coldatoms}
I. Bloch, J. Dalibard, and W. Zwerger, \emph{Many-body physics with ultracold gases}, \href{https://doi.org/10.1103/RevModPhys.80.885}{Rev. Mod. Phys. {\bf 80}, 885 (2008)}.

\bibitem{Newtoncradle}
T. Kinoshita, T. Wenger, and D. S. Weiss, \emph{A quantum Newton's cradle}, \href{https://doi.org/10.1038/nature04693}{Nature {\bf 440}, 900 (2006)}.

\bibitem{sbdd-20}
M. Schemmer, I. Bouchoule, B. Doyon, and J. Dubail, {\it Generalized HydroDynamics on an Atom Chip},
\href{https://doi.org/10.1103/PhysRevLett.122.090601}{Phys. Rev. Lett. {\bf 122}, 090601 (2019)}

\bibitem{langen-2015}
T.~Langen, S.~Erne, R.~Geiger, B.~Rauer, T.~Schweigier, M.~Kuhnert, W.~Rohringer, I.~E.~Mazets, T.~Gasenzer, J.~Schmiedmayer, 
Experimental observation of a generalized Gibbs ensemble,
\href{http://dx.doi.org/10.1126/science.1257026}{Science {\bf 348}, 207 (2015)}.

\bibitem{BKP:dualunitary}
B. Bertini, P. Kos, and T. Prosen,
\emph{Exact Correlation Functions for Dual-Unitary Lattice Models in $1+1$ Dimensions}, \href{https://doi.org/10.1103/PhysRevLett.123.210601}{Phys. Rev. Lett. {\bf 123}, 210601 (2019)}. 







\bibitem{SK:EOM}
M. Stark and M. Kollar, 
\textit{Kinetic description of thermalization dynamics in weakly interacting quantum systems}, 
\href{http://arxiv.org/abs/1308.1610}{arXiv:1308.1610 (2013)}. 


\bibitem{NessiArxiv15}
N. Nessi and A. Iucci, 
\textit{Glass-like Behavior in a System of One Dimensional Fermions after a Quantum Quench}, 
\href{http://arxiv.org/abs/1503.02507}{arXiv:1503.02507 (2015)}.

\bibitem{INJPhys}
 A. Iucci and N. Nessi, 
 \textit{Equations of Motion for the Out-of-Equilibrium Dynamics of Isolated Quantum Systems from the Projection Operator Technique}, 
 \href{http://dx.doi.org/10.1088/1742-6596/568/1/012013}{J. Phys.: Conf. Ser. {\bf 568}, 012013 (2014)}.

\bibitem{ESY:QBE}
L. Erd\H{o}s, M. Salmhofer, and H.-T. Yau, 
\textit{On the Quantum Boltzmann Equation}, 
\href{http://dx.doi.org/10.1023/B:JOSS.0000037224.56191.ed}{J. Stat. Phys. {\bf 116}, 367 (2004)}.


\bibitem{LS:QBE}
J. Lukkarinen and H. Spohn,
\textit{Not to Normal Order--Notes on the Kinetic Limit for Weakly Interacting Quantum Fluids}, 
\href{http://dx.doi.org/10.1007/s10955-009-9682-8}{J. Stat. Phys. {\bf 134}, 1133 (2009)}.

\bibitem{FMS:QBE}
M. L. R. F\"urst, C. B. Mendl, and H. Spohn, 
\textit{Matrix-valued Boltzmann equation for the Hubbard chain}, 
\href{http://dx.doi.org/10.1103/PhysRevE.86.031122}{Phys. Rev. E {\bf 86}, 031122 (2012)}.

\bibitem{FMS:NIQBE}
M. L. R. F\"urst, C. B. Mendl, and H. Spohn, 
\textit{Matrix-valued Boltzmann equation for the nonintegrable Hubbard chain}, 
\href{http://dx.doi.org/10.1103/PhysRevE.88.012108}{Phys. Rev. E {\bf 88}, 012108 (2013)}.

\bibitem{KBbook}
L.~P.~Kadanoff and G.~A.~Baym, \emph{Quantum statistical mechanics} (Benjamin, New York, 1962).

\bibitem{Bonitzbook}
M. Bonitz, \emph{Quantum Kinetic Theory} (Teubner, Stuttgart, 1998).








\bibitem{Orbach}
R. Orbach, 
\textit{Linear Antiferromagnetic Chain with Anisotropic Coupling}, 
\href{http://dx.doi.org/10.1103/PhysRev.112.309}{Phys. Rev. {\bf 112}, 309 (1958)}.

\bibitem{EKreview}
F. H. L. Essler and R. M. Konik in 
\textit{From Fields to Strings: Circumnavigating Theoretical Physics},
edited by M. Shifman, A. Vainshtein, and J. Wheater (World Scientific, Singapore, 2005); 
\textit{Applications of Massive Integrable Quantum Field Theories to Problems in Condensed Matter Physics}, 
\href{http://arxiv.org/abs/cond-mat/0412421}{arXiv:0412421 (2004)}.


\bibitem{EF16}
F.~H.~L. Essler and M. Fagotti, 
\textit{Quench dynamics and relaxation in isolated integrable quantum spin chains}, 
\href{http://iopscience.iop.org/article/10.1088/1742-5468/2016/06/064002}{J. Stat. Mech. (2016) 064002}.


\bibitem{sc-14}
S. Sotiriadis and P. Calabrese, {\it Validity of the GGE for quantum quenches from interacting to noninteracting models},
\href{http://dx.doi.org/10.1088/1742-5468/2014/07/P07024}{J. Stat. Mech. (2014) P07024}.

\bibitem{sps-04}
K. Sengupta, S. Powell, S. Sachdev, 
Quench dynamics across quantum critical points, 
\href{https://doi.org/10.1103/PhysRevA.69.053616}{Phys. Rev. A {\bf 69}, 053616 (2004)}.

\bibitem{CEF1}
P.~Calabrese, F.~H.~L.~Essler, and M.~Fagotti,
\textit{Quantum Quench in the Transverse Field Ising chain I: Time evolution of order parameter correlators}, 
\href{http://dx.doi.org/10.1088/1742-5468/2012/07/P07016}{J. Stat. Mech. (2012) P07016}. 

\bibitem{SMT}
S. Sotiriadis, G. Takacs, and G. Mussardo, 
\emph{Boundary State in an Integrable Quantum Field Theory Out of Equilibrium},
\href{http://dx.doi.org/10.1016/j.physletb.2014.04.058}{Phys. Lett. B {\bf 734}, 52 (2014)}.



\bibitem{BrunoThesis}
B. Bertini, \emph{Non-equilibrium dynamics of interacting many-body quantum systems in one
dimension}, \href{https://ora.ox.ac.uk/objects/uuid:1e2c50b9-73b3-4ca0-a5f3-276f967c3720}{PhD Thesis, University of Oxford, (2015)}. 



\bibitem{W:CUT} 
F. Wegner, 
\textit{Flow-equations for Hamiltonians}, 
\href{http://dx.doi.org/10.1002/andp.19945060203}{Ann. Physik {\bf 506}, 77 (1994)}.

\bibitem{UhrigCUT}
C. Knetter and G. S. Uhrig, 
\textit{Perturbation theory by flow equations: dimerized and frustrated S = 1/2 chain}, 
\href{http://dx.doi.org/10.1007/s100510050026}{Eur. Phys. J. B {\bf 13}, 209 (2000)}.

\bibitem{UhrigCUT2}
C. P. Heidbrink and G. S. Uhrig, 
\textit{Renormalization by continuous unitary transformations: one-dimensional spinless fermions}, 
\href{http://dx.doi.org/10.1140/epjb/e2002-00401-9}{Eur. Phys. J. B {\bf 30}, 443 (2002)}.

\bibitem{kehreinbook}
S. Kehrein, 
\textit{The flow-equation approach to many-particle systems} (Springer, Berlin, 2007).


\bibitem{Shivamoggibook}
B. K. Shivamoggi, \emph{Perturbation Methods for Differential Equations} (Birkh\"auser, Boston, 2003). 

\bibitem{LMMR:14}
J. Lux, J. M\"uller, A. Mitra, and A. Rosch, \emph{Hydrodynamic long- time tails after a quantum quench}, \href{https://doi.org/10.1103/PhysRevA.89.053608}{Phys. Rev. A {\bf 89}, 053608 (2014)}.

\bibitem{KBCHH:15}
H. Kim, M. C. Ba{\~n}uls, J. I. Cirac, M. B. Hastings, and D. A. Huse, \emph{Slowest local operators in quantum spin chains}, \href{https://doi.org/10.1103/PhysRevE.92.012128}{Phys. Rev. E {\bf 92,} 012128 (2015)}.





\bibitem{peschel2003}
I. Peschel, {\it Calculation of reduced density matrices from correlation functions},
\href{http://dx.doi.org/10.1088/0305-4470/36/14/101}{J. Phys. A {\bf 36}, L205 (2003).}

\bibitem{lrv-03}
J. I. Latorre, E. Rico, and G. Vidal,
{\it Ground state entanglement in quantum spin chains},
\href{}{Quant. Inf. Comp. {\bf 4}, 048 (2004)}.

\bibitem{pe-09}
I. Peschel and V. Eisler, {\it Reduced density matrices and entanglement entropy in free lattice models}, 
\href{https://iopscience.iop.org/article/10.1088/1751-8113/42/50/504003} { J. Phys. A {\bf 42}, 504003 (2009)}.

\bibitem{BK:rates}
F.~R.~A.~Biebl and S.~Kehrein,
\emph{Thermalization rates in the one dimensional Hubbard model with next-to-nearest neighbor hopping}, 
\href{http://dx.doi.org/10.1103/PhysRevB.95.104304}{Phys. Rev. B {\bf 95}, 104304 (2017)}.

\bibitem{ac-18b}
V. Alba and P. Calabrese, {\it Quantum information dynamics in multipartite integrable systems},  
\href{https://doi.org/10.1209/0295-5075/126/60001}{EPL {\bf 126}, 60001 (2019)}

\bibitem{ac-17a}
V. Alba and P. Calabrese, {\it Quench action and R\'enyi entropies in integrable systems},
\href{https://doi.org/10.1103/PhysRevB.96.115421}{Phys. Rev. B {\bf 96}, 115421 (2017)}.

\bibitem{ac-17b}
V. Alba and P. Calabrese, {\it R\'enyi entropies after releasing the N\'eel state in the XXZ spin-chain}, 
\href{http://dx.doi.org/10.1088/1742-5468/aa934c}{J. Stat. Mech. (2017) 113105}.

\bibitem{mac-18}
M. Mestyan, V. Alba, and P. Calabrese, {\it R\'enyi entropies of generic thermodynamic macrostates in integrable systems},
\href{https://doi.org/10.1088/1742-5468/aad6b9}{J. Stat. Mech. (2018) 083104}




\end{thebibliography}
\end{document}